\documentclass[preprint,12pt,authoryear]{elsarticle}

\usepackage{amssymb}
\usepackage{amsmath}
\usepackage{graphicx}
\usepackage{subcaption}
\usepackage{tikz}
\usetikzlibrary{arrows.meta,positioning,shapes.geometric,fit}
\usepackage{lineno}
\usepackage{hyperref}

\journal{Urban Climate}

\begin{document}

\begin{frontmatter}



\title{Fractal-based variable-drag model for porous-media tree representations} 


\author[inst1]{Takumi Tokiwa\corref{cor1}}
\ead{tokiwa.t.3c84@m.isct.ac.jp}
\author[inst1]{Yuwei Yin}
\ead{yin.y.1b8b@m.isct.ac.jp}
\author[inst2]{Ryo Onishi}
\ead{onishi.ryo@scrc.iir.isct.ac.jp}

\cortext[cor1]{Corresponding author.}

\affiliation[inst1]{
  organization={Faculty of Engineering, Institute of Science Tokyo},
  city={Tokyo},
  postcode={152-8550},
  country={Japan}
}
\affiliation[inst2]{
  organization={Supercomputing Research Center, Institute of Integrated Research, Institute of Science Tokyo},
  city={Tokyo},
  postcode={152-8550},
  country={Japan}
}

\begin{abstract}
Accurate representation of trees is essential for predictive urban micrometeorological simulations, but explicitly resolving detailed tree geometry is computationally prohibitive. Trees are therefore commonly represented as porous media, often with a constant drag coefficient even when spatially heterogeneous area-density distributions are introduced. This limits transferability across inflow conditions and can increase sensitivity to grid resolution, particularly in the grayscale regime where a tree is represented by only a few computational cells.

We propose a fractal-based variable-drag model for porous-media tree representations, in which the drag coefficient is prescribed cell-wise as $C_D=C_D(n_{\mathrm{eff}},Re_{\mathrm{eff}})$. Here, $n_{\mathrm{eff}}$ is the cell-effective branching order representing unresolved local morphological complexity, and $Re_{\mathrm{eff}}$ is the cell-effective Reynolds number representing the local flow regime. The model is assessed using steady Reynolds-averaged Navier--Stokes simulations of a porous fractal tree over systematic sweeps of grid resolution and inflow velocity. Model performance is evaluated primarily using aerodynamic porosity, which measures the bulk momentum attenuation induced by the tree.

The proposed model produces a plausible aerodynamic response and improves robustness to grid resolution compared with constant-$C_D$ models. It also captures the variation of bulk drag across inflow conditions without empirical retuning. Notably, this whole-tree response is recovered through local cell-wise quantities, $n_{\mathrm{eff}}$ and $Re_{\mathrm{eff}}$. These results demonstrate that morphology- and flow-dependent drag provides a practical route to improving porous-media tree modeling.
\end{abstract}




\end{frontmatter}



\section{Introduction}
In recent years, accurate urban micrometeorological forecasting has become increasingly important for urban environmental assessment and planning, particularly in the context of accelerating urban population growth, intensifying heat-related risks, and the development of smart societies~\citep{UN:WUP2019,IPCC2023,LancetCountdown2024,UNHabitat2024}. Meeting this demand requires physically credible representations of the processes that shape the urban atmosphere, including the effects of vegetation.

Trees play a vital role in regulating urban micrometeorology by modulating wind flow, temperature distribution, and pollutant dispersion~\citep{GromkeRuck2009,Li2024Review,Reznicek2025}. These impacts are governed by aerodynamic interactions between the flow and the multi-scale architecture of trees, through which trees extract momentum and alter turbulence. Consequently, a reliable description of tree-induced aerodynamic resistance is a prerequisite for predictive simulations of urban ventilation, scalar transport, and related microclimatic conditions.

For practical urban-scale computations, however, resolving detailed tree geometry is rarely feasible. Tree representations in CFD can be viewed as a resolution-dependent hierarchy: (i) explicit simulations resolve branch and trunk geometry and offer high-fidelity reference solutions, albeit at high computational cost~\citep{Yin2025,Yin2025PoF}; (ii) implicit approaches parameterize vegetation effects as fully subgrid forcing~\citep{BrownWilliams1998,LoughnerEtAl2012}; and (iii) an increasingly common intermediate regime represents a tree with only a few grid cells, referred to here as grayscale simulations~\citep{Oshio2021,Gromke2015a,Gromke2015b}. As computational resources improve, grayscale simulations are becoming more relevant for real-world applications, where fully resolved simulations remain too expensive while fully homogenized single-cell representations are overly restrictive.

In both Reynolds-averaged Navier--Stokes (RANS) and large-eddy simulation (LES) frameworks, implicit and grayscale approaches typically model trees as porous media, i.e., distributed momentum sinks. In such models, the canopy effect is commonly expressed through a drag term proportional to a local area density and a drag coefficient, e.g., $\mathbf{S}_u \propto -C_D \alpha |\mathbf{u}|\mathbf{u}$, where $\alpha(\mathbf{x})$ denotes an area-density field such as plant or leaf area density (PAD/LAD). A common conventional practice is to prescribe spatially uniform area density over the tree cells together with a constant drag coefficient $C_D$, or to tune a constant $C_D$ on a case-by-case basis, as in baseline porous-canopy implementations~\citep{Matsuda2018,Gromke2015a,Gromke2015b}. More recently, advanced conventional models have incorporated grid- or voxel-resolved PAD/LAD distributions derived from geometry or measurements to better represent spatial heterogeneity~\citep{Oshio2021,Zhang2026EACFM,Rodriguez2024}, while still retaining a constant $C_D$. These developments have improved the geometric realism of porous-tree representations. However, the drag coefficient itself is still typically prescribed as a constant and therefore does not explicitly respond to local flow regime or unresolved morphological complexity within each cell. This remaining limitation can reduce transferability across inflow conditions~\citep{Manickathan2018,Bekkers2022} and can introduce sensitivity to grid resolution~\citep{Fu2024,Zhang2026EACFM}, particularly in the grayscale regime where the effective unresolved morphology represented within each cell changes with cell size.

These issues motivate a drag formulation in which the cell-wise drag coefficient adapts to both morphology and flow regime. In this study, we propose a fractal-based variable-drag model for porous-media tree representations in which the drag coefficient is prescribed as $C_D=C_D(n_{\mathrm{eff}},Re_{\mathrm{eff}})$. Here, $n_{\mathrm{eff}}$ is the cell-effective branching order
representing unresolved local morphological complexity, and
$Re_{\mathrm{eff}}$ is the cell-effective Reynolds number
characterizing the local flow regime.

We validate the proposed model using steady RANS simulations with systematic parameter sweeps in both grid resolution and inflow velocity. To isolate the bulk momentum attenuation attributable to the canopy forcing, we evaluate model performance primarily using aerodynamic porosity~\citep{Guan2003,Bitog2011}, defined as the ratio of leeward to windward plane-averaged streamwise velocity across the tree. We compare the proposed model against (i) a general conventional baseline using uniform PAD with constant $C_D$ and (ii) an advanced conventional baseline using voxel-resolved PAD with constant $C_D$.

The remainder of this paper is organized as follows. In Sec.~\ref{sec:drag_model}, we present the fractal-based variable-drag formulation.
We first describe the dependence of $C_D$ on branching order and Reynolds number, which represent morphological complexity and flow regime, respectively. We then define $n_{\mathrm{eff}}$ and $Re_{\mathrm{eff}}$ for its cell-wise evaluation. In Sec.~\ref{sec:simulation setup}, we describe the RANS methodology and computational setup, summarize the grid-resolution and inflow-velocity sweeps, and define the aerodynamic porosity metric used for evaluation. In Sec.~\ref{sec:RaD}, we present the results and discussion, focusing on qualitative flow behavior, resolution robustness, and Reynolds-number dependence compared with constant-$C_D$ models. Finally, Sec.~\ref{sec:conclusion} concludes the paper and outlines future work.

\section{Fractal-based variable-drag formulation}
\label{sec:drag_model}

\subsection{Dependence of $C_D$ on branching order and Reynolds number}

The proposed variable-drag model accounts for the dependence of the
drag coefficient on morphological complexity and flow regime. For
complete fractal trees, morphological complexity is parameterized by
the branching order $n$, which denotes the order of recursive
branching in the tree construction. Here, $n=1$ corresponds to a
single cylinder, and increasing $n$ introduces progressively finer
branches. The flow regime is characterized by $Re_H$, the Reynolds
number based on tree height. The drag-coefficient dependence for
complete fractal trees is therefore represented by $C_D(n,Re_H)$.

\begin{figure}[htbp]
\centering
\includegraphics[width=\linewidth]{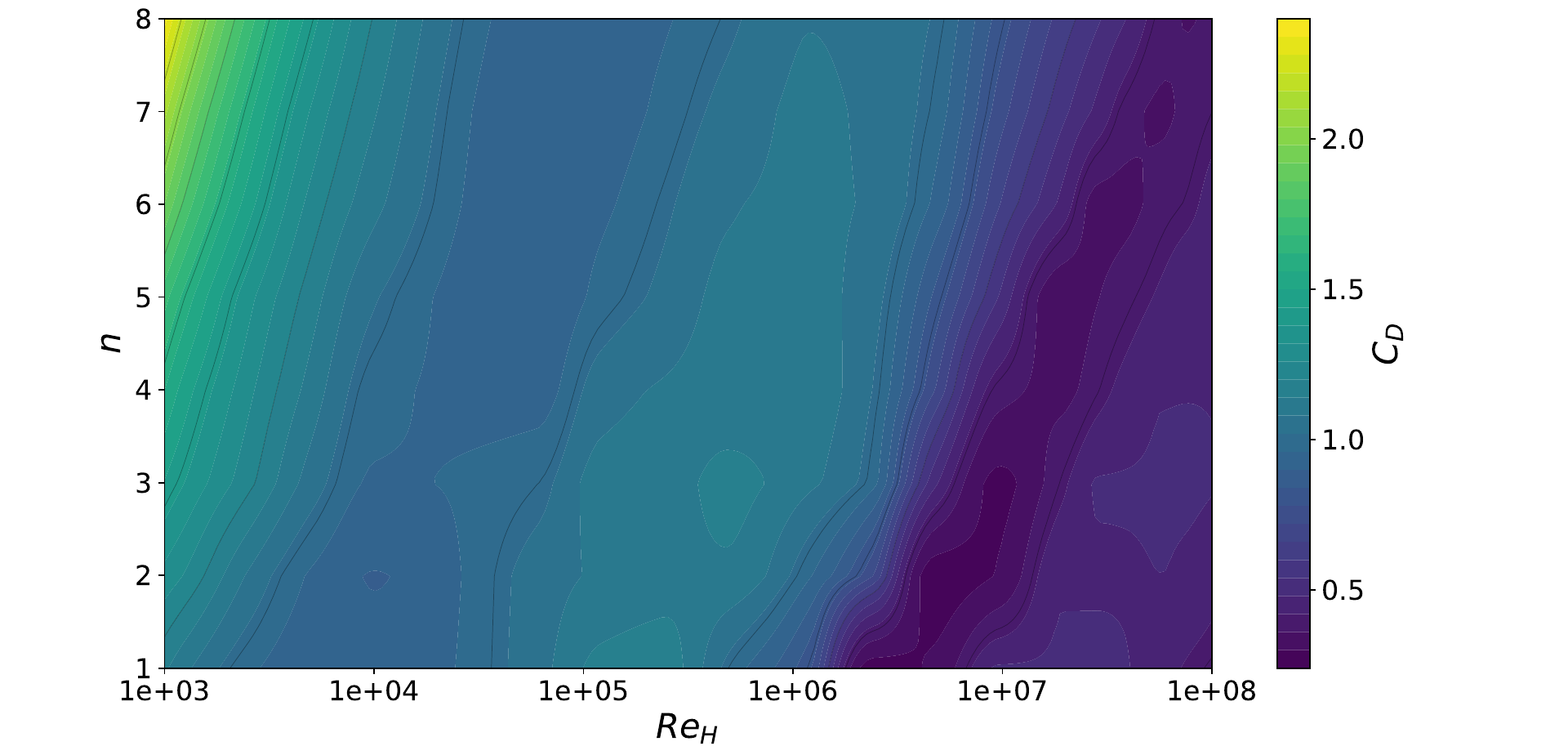}
\caption{\label{fig:cd_table}
A priori relationship $C_D(n,Re_H)$ for Tree~A.}
\end{figure}

We construct the a priori relationship $C_D(n,Re_H)$ using our
analytical drag-estimation method for fractal trees
~\citep{Tokiwa2026}. The detailed derivation of this method is
provided by \citet{Tokiwa2026}. In the present study, the relationship is
constructed for Tree~A, which is the target geometry used in the
subsequent RANS simulations, and stored as a lookup table for
interpolation. Figure~\ref{fig:cd_table} presents the resulting
relationship between branching order, tree-height-based Reynolds
number, and drag coefficient.

The a priori relationship $C_D(n,Re_H)$ is constructed for complete
fractal trees, whereas the drag coefficient in the porous-media model
is evaluated for individual computational cells. Its cell-wise application therefore requires corresponding
cell-effective measures of morphological complexity and flow regime,
denoted by $n_{\mathrm{eff}}$ and $Re_{\mathrm{eff}}$,
respectively.

\subsection{Cell-wise descriptors:
$n_{\mathrm{eff}}$ and $Re_{\mathrm{eff}}$}

\subsubsection{Effective branching order from fractal self-similarity.}

\begin{figure}[htbp]
\centering
\includegraphics[width=\linewidth]{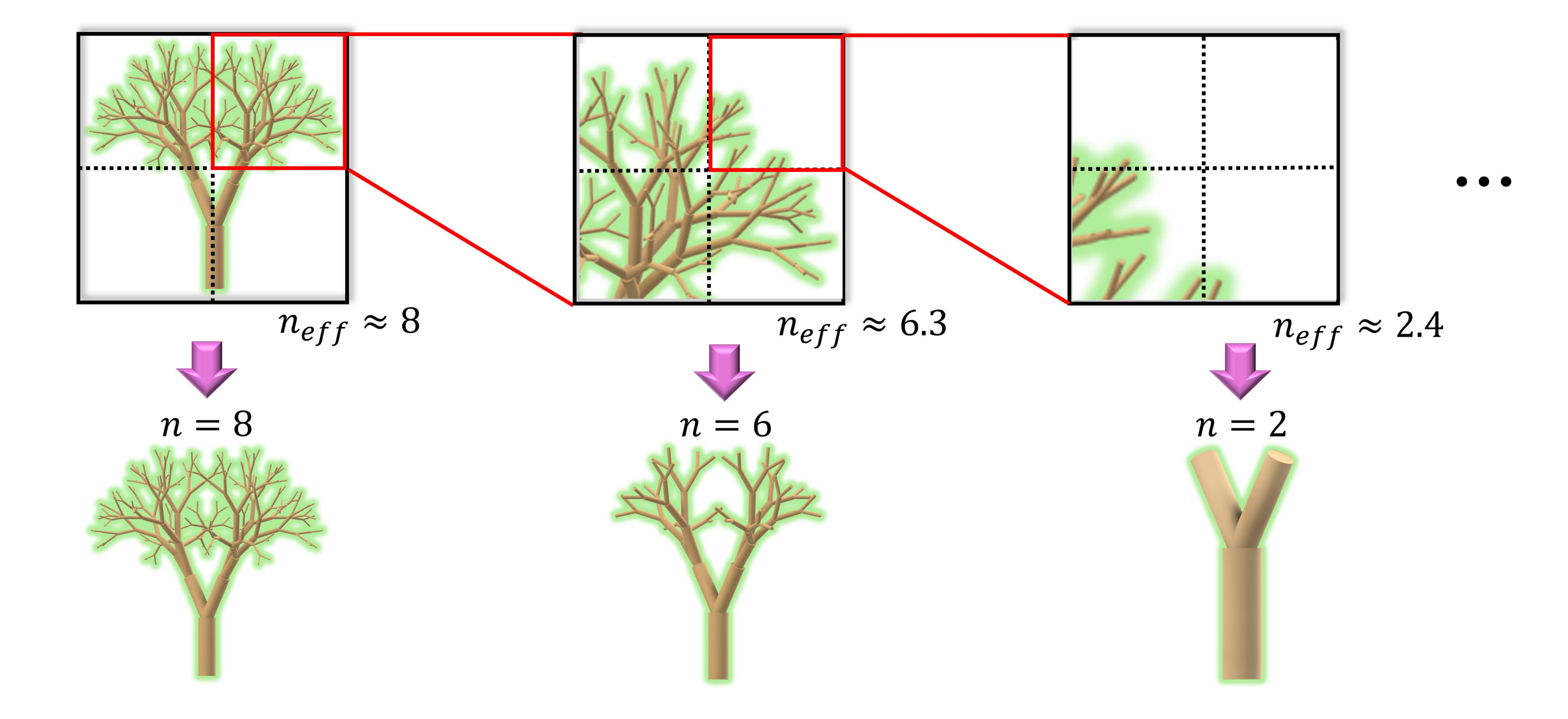}
\caption{\label{fig:grayscale_schematic}
Schematic illustration of grid-dependent in-cell morphology and its
interpretation via fractal self-similarity.}
\end{figure}

The morphology represented within each computational cell changes
with grid resolution, as illustrated in
Fig.~\ref{fig:grayscale_schematic}. Because the present fractal trees
have a self-similar branching structure, each in-cell geometry can be
associated with a complete reference tree of equivalent morphological
complexity. We define the branching order of this equivalent reference
tree as the cell-effective branching order $n_{\mathrm{eff}}$, which
characterizes the unresolved, grid-dependent morphological complexity
within the cell.

To infer $n_{\mathrm{eff}}$, we define the dimensionless cell-wise
morphology index
$\Omega(\mathbf{x})=\mathrm{SVR}\times R_g$.
Here, $\mathrm{SVR}$ is the surface-to-volume ratio within a cell,
$\mathrm{SVR}=(\text{surface area})/
(\text{tree geometry volume})$ $[\mathrm{m}^{-1}]$, and
$R_g$ $[\mathrm{m}]$ is the radius of gyration of the in-cell
geometry, defined as
$R_g=\sqrt{\left(\sum_i V_i
\left|\mathbf{c}_i-\mathbf{R}_{\mathrm{com}}\right|^2\right)
\big/\left(\sum_i V_i\right)}$.
In this expression, $V_i$ and $\mathbf{c}_i$ are the volume and
centroid of the $i$th constituent element, respectively, and
$\mathbf{R}_{\mathrm{com}}
=\left(\sum_i V_i\mathbf{c}_i\right)/
\left(\sum_i V_i\right)$
is the volume-weighted center position. The radius of gyration
represents the statistical spread of the constituent volume elements
around this center position.

\begin{figure}[htbp]
\centering
\begin{subfigure}[t]{0.43\linewidth}
  \centering
  \includegraphics[width=\linewidth]{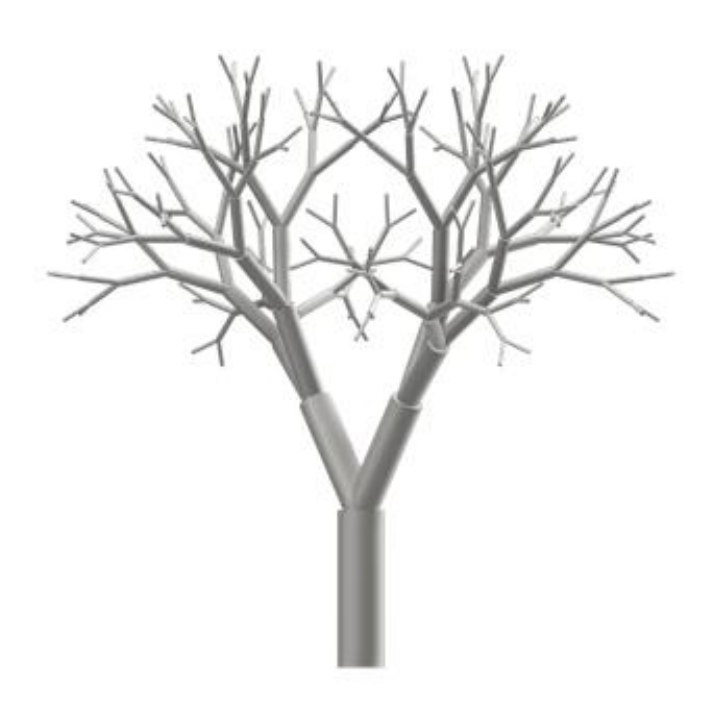}
  \caption{Tree A: geometry}
\end{subfigure}\hspace{0.03\linewidth}
\begin{subfigure}[t]{0.50\linewidth}
  \centering
  \includegraphics[width=\linewidth]{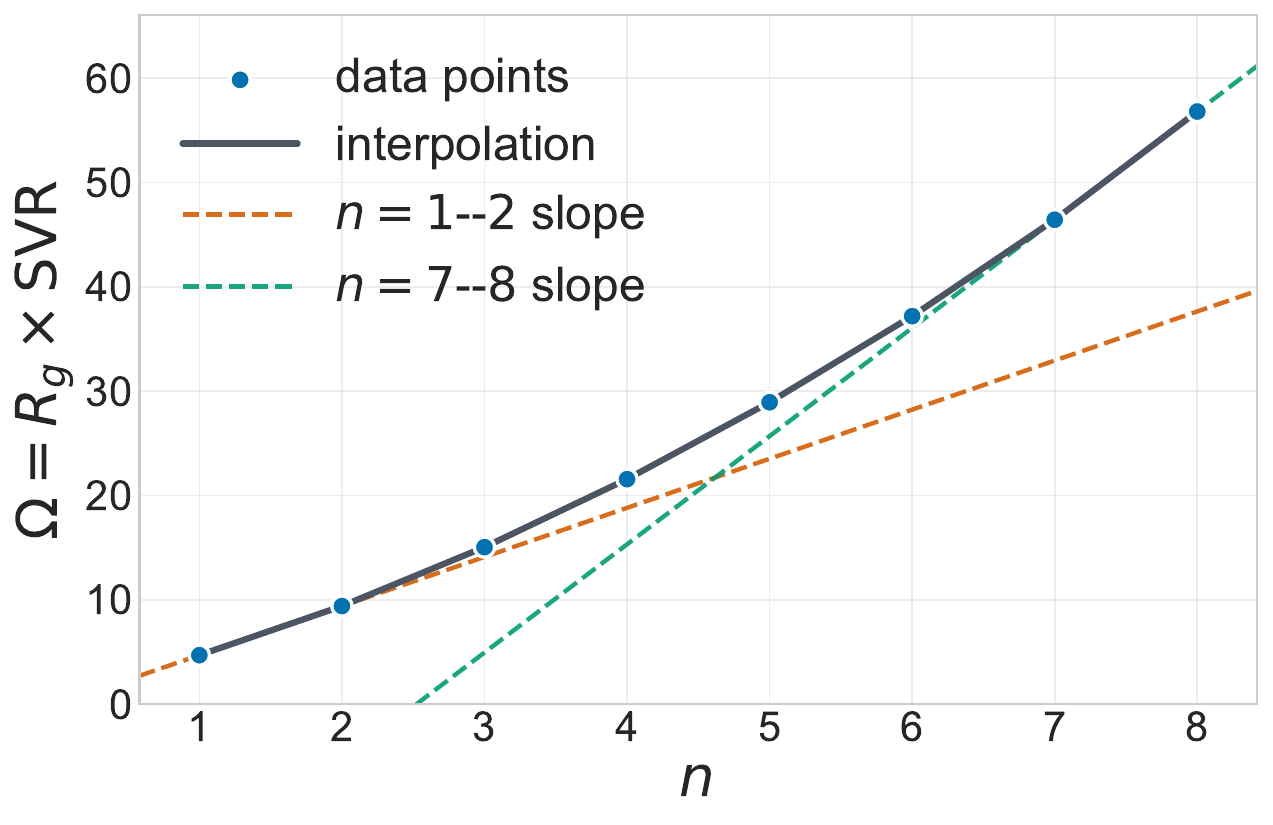}
  \caption{Tree A: $\Omega$--$n$}
\end{subfigure}

\vspace{0.8ex}

\begin{subfigure}[t]{0.45\linewidth}
  \centering
  \includegraphics[width=\linewidth]{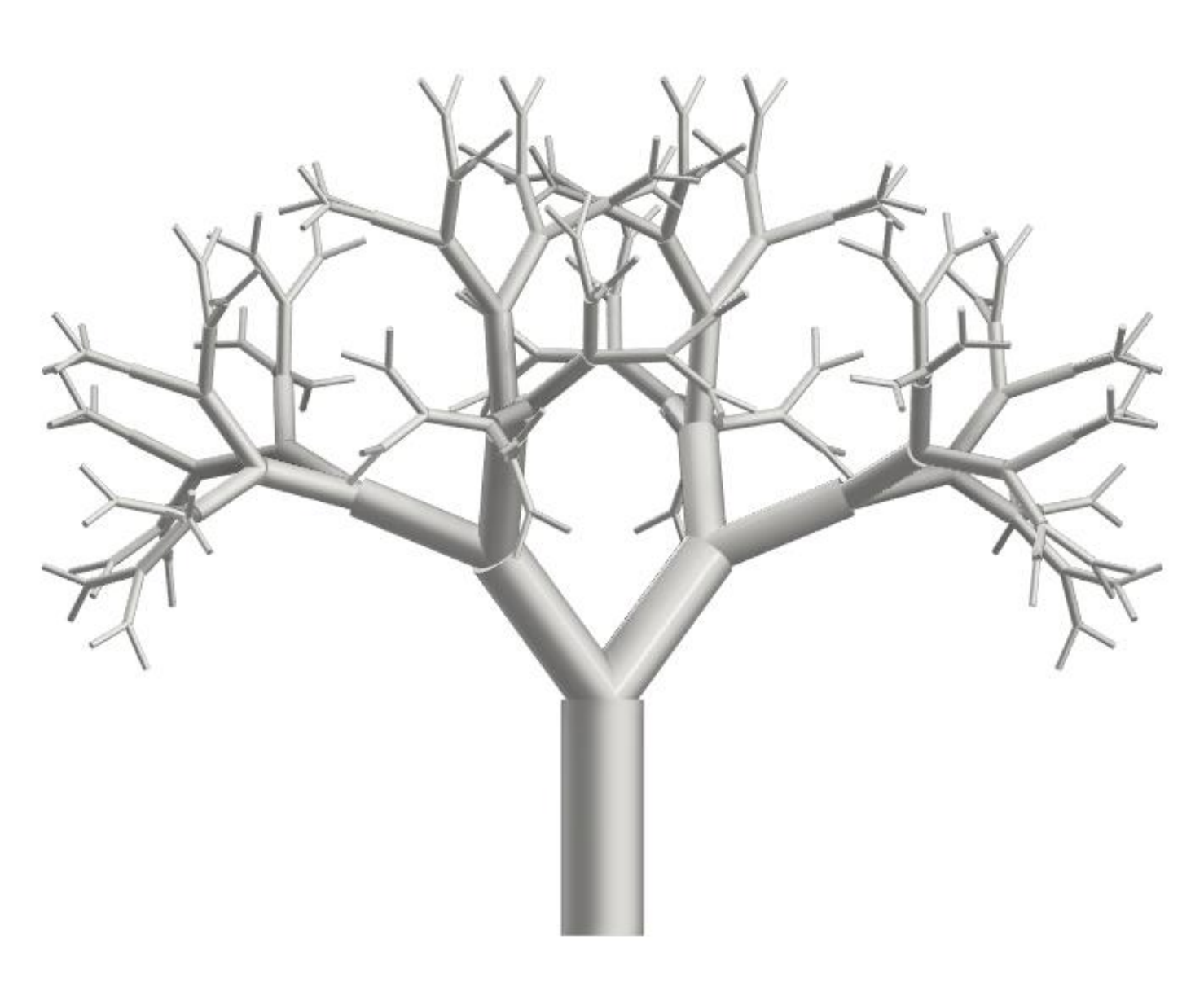}
  \caption{Tree B: geometry}
\end{subfigure}\hspace{0.03\linewidth}
\begin{subfigure}[t]{0.50\linewidth}
  \centering
  \includegraphics[width=\linewidth]{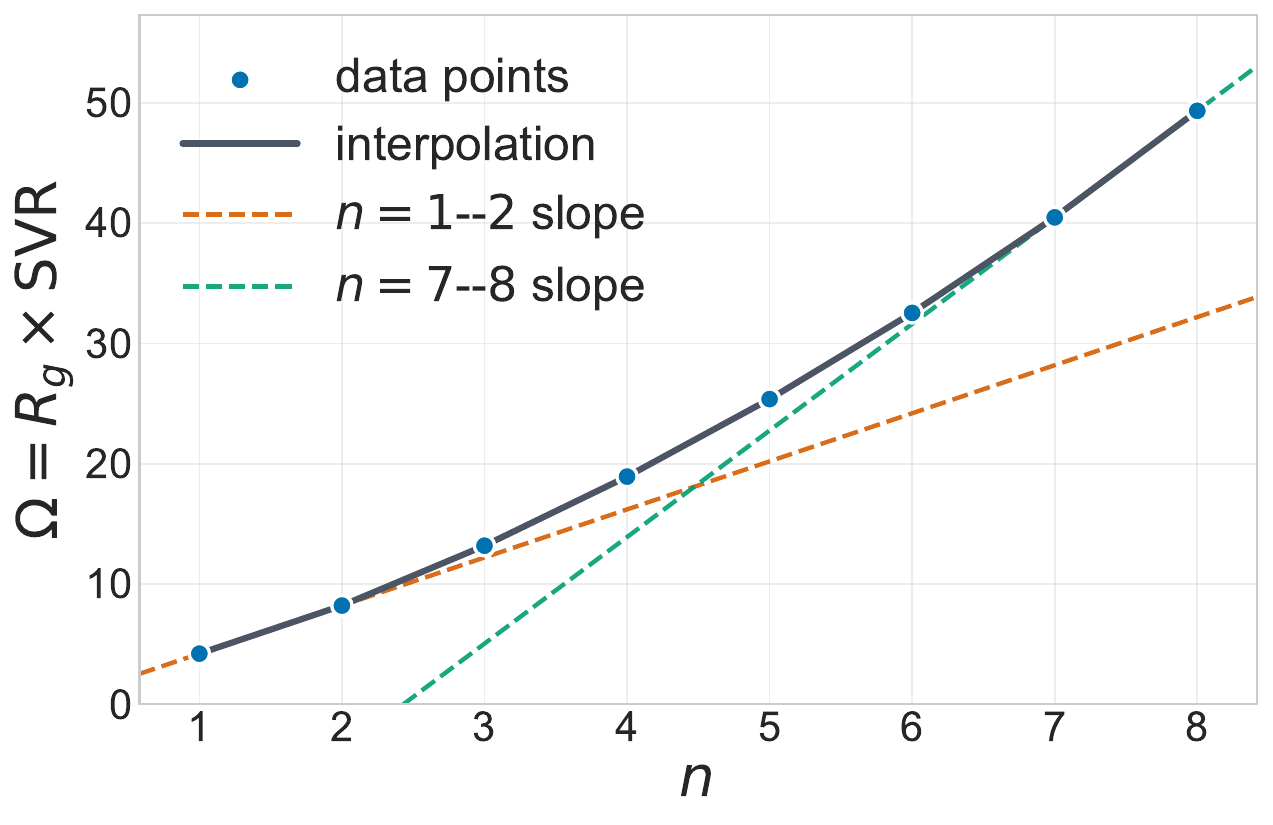}
  \caption{Tree B: $\Omega$--$n$}
\end{subfigure}

\vspace{0.8ex}

\begin{subfigure}[t]{0.40\linewidth}
  \centering
  \includegraphics[width=\linewidth]{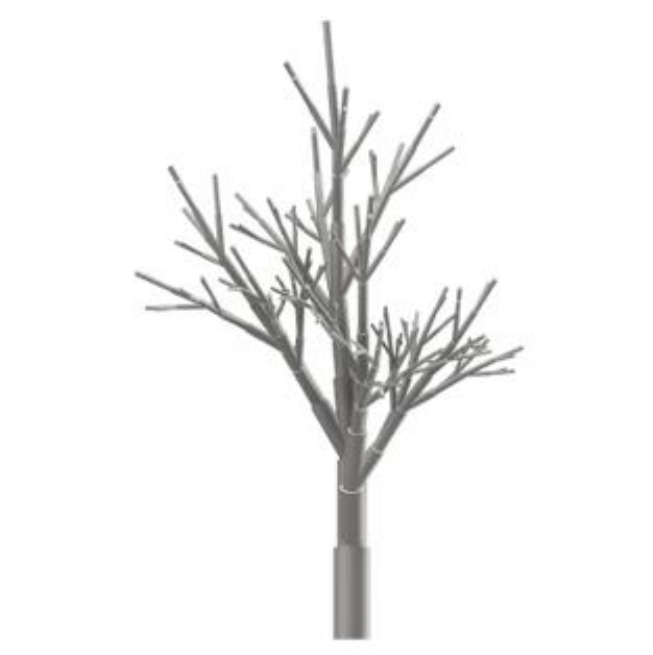}
  \caption{Tree C: geometry}
\end{subfigure}\hspace{0.03\linewidth}
\begin{subfigure}[t]{0.50\linewidth}
  \centering
  \includegraphics[width=\linewidth]{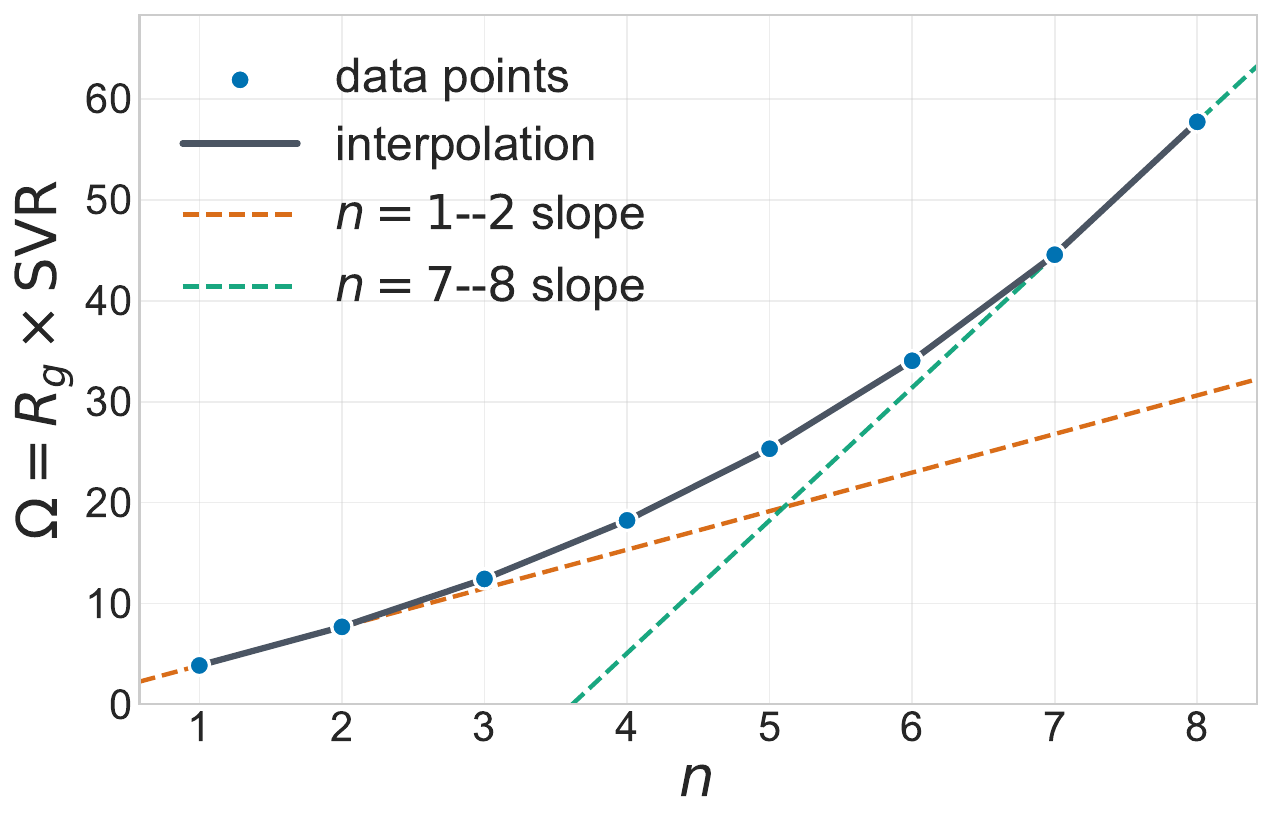}
  \caption{Tree C: $\Omega$--$n$}
\end{subfigure}

\caption{\label{fig:trees_omega_n}
Fractal-tree geometries and corresponding reference
$\Omega$--$n$ relationships for Trees A--C.}
\end{figure}

For each fractal-tree family, we construct a reference
$\Omega\!-\!n$ relationship by evaluating $\Omega$ for complete trees
at discrete branching orders. Because $n_{\mathrm{eff}}=n$ for a
complete reference tree, the cell-effective branching order
$n_{\mathrm{eff}}(\mathbf{x})$ can be inferred by applying this
reference relationship to the cell-wise value $\Omega(\mathbf{x})$
through lookup and interpolation. We demonstrate this procedure using
three fractal-tree families, Trees A, B, and C, generated with an
L-system algorithm~\citep{Prusinkiewicz1996}. For each tree family, the
reference relationship is constructed from complete trees with
branching orders $n=1,\dots,8$.
Figure~\ref{fig:trees_omega_n} presents the $n=8$ geometries and the
corresponding $\Omega\!-\!n$ relationships for Trees A--C.

As shown in Fig.~\ref{fig:trees_omega_n}, $\Omega$ increases
monotonically with $n$ for all three tree families, indicating that it
can be used as an index of morphological complexity for the present
fractal-tree families. Because all reference trees are normalized to
a height of $H=1\,\mathrm{m}$, the scale of the radius of gyration
$R_g$ is constrained by $H$. Therefore, the dependence of $\Omega$ on
$n$ can be interpreted as primarily reflecting the variation in
$\mathrm{SVR}$. The plots also include reference lines based on the
local growth rates at the lowest branching orders, from $n=1$ to $2$,
and the highest branching orders, from $n=7$ to $8$. The lower-order
reference line characterizes the trend in geometries dominated by the
trunk and thick branches, whereas the higher-order reference line
characterizes the trend in structures strongly influenced by fine
branches. The steeper slope of the higher-order reference line is
consistent with the geometric nature of fractal trees: the number of
branches multiplies with each generation, and the resulting fine
branches provide a large surface area relative to their volume,
increasing their contribution to $\Omega$.

\begin{figure}[htbp]
\centering
\includegraphics[width=0.9\linewidth]{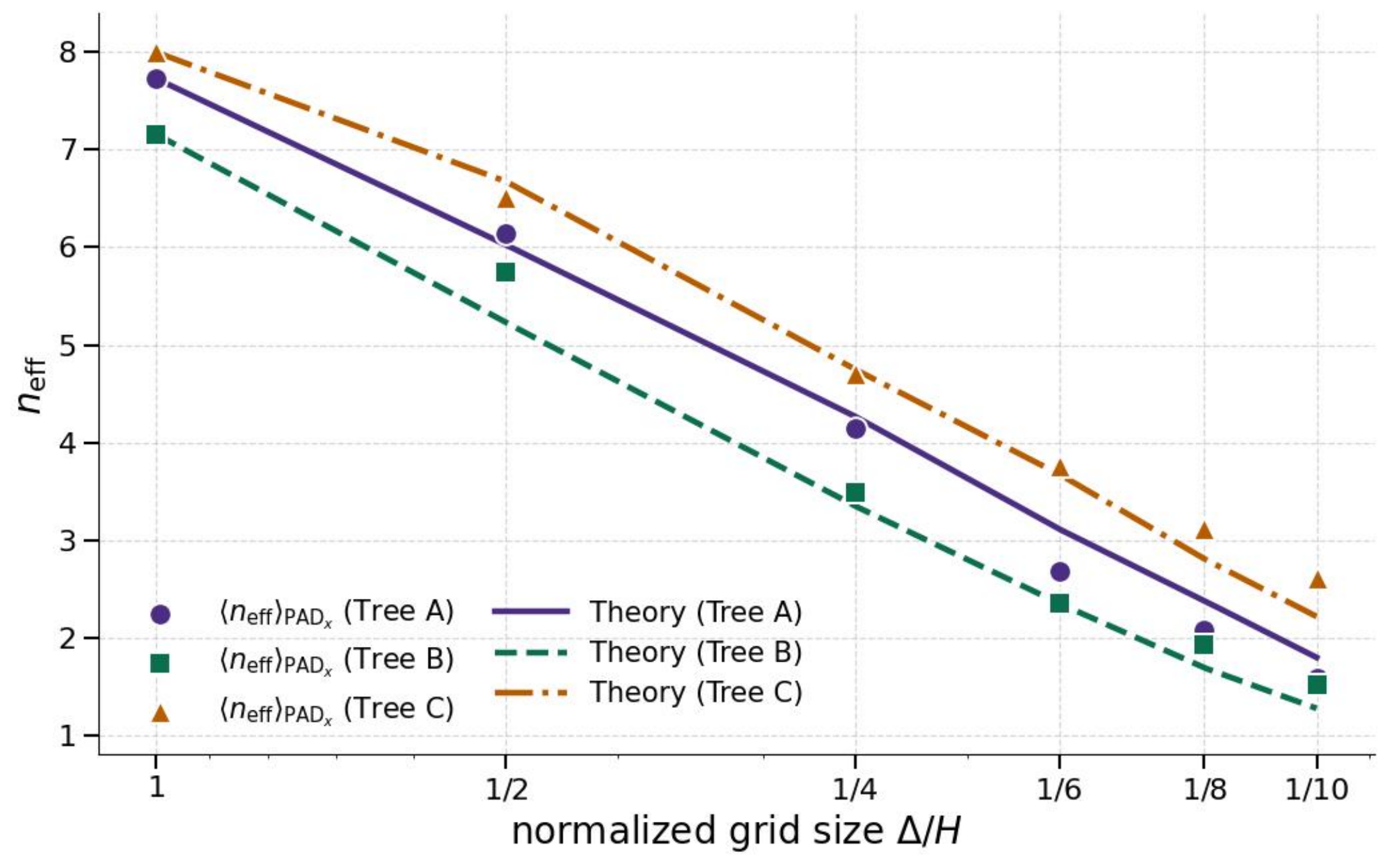}
\caption{\label{fig:neff_validation}
Validation of the $n_{\mathrm{eff}}$ inference across grid resolution.}
\end{figure}

To evaluate the inferred $n_{\mathrm{eff}}$ across grid resolutions,
we compare its $\mathrm{PAD}_x$-weighted mean,
$\langle n_{\mathrm{eff}}\rangle_{\mathrm{PAD}_x}$, with the
theoretical prediction $n_{\mathrm{eff}}^{\mathrm{th}}(\Delta)$
derived in~\ref{app:neff_prediction}.
Here, $\mathrm{PAD}_i$ denotes the projected area of tree elements
contained in a computational cell divided by the cell volume in the
$i$th direction, and $\mathrm{PAD}_x$ is its streamwise component,
with $x$ corresponding to the inflow direction in the subsequent
RANS simulations.
Figure~\ref{fig:neff_validation} presents the comparison for
$\Delta/H\in\{1,\,1/2,\,1/4,\,1/6,\,1/8,\,1/10\}$.
For all three families,
$\langle n_{\mathrm{eff}}\rangle_{\mathrm{PAD}_x}$ follows the
grid-dependent trend predicted by
$n_{\mathrm{eff}}^{\mathrm{th}}$, supporting the proposed inference
method.

\subsubsection{Effective Reynolds number.}

As the cell-wise counterpart of $Re_H$, we define the cell-effective
Reynolds number as
$Re_{\mathrm{eff}}(\mathbf{x})
=|\mathbf{u}(\mathbf{x})|L_{\mathrm{cell}}(\mathbf{x})/\nu$,
where $\mathbf{u}(\mathbf{x})$ is the local velocity, $\nu$ is the
kinematic viscosity, and $L_{\mathrm{cell}}(\mathbf{x})$ is a
morphology-based characteristic length.

In this study, we determine $L_{\mathrm{cell}}(\mathbf{x})$ from the
in-cell surface area $S_{\mathrm{cell}}(\mathbf{x})$ to maintain the
scale relationship between the tree height and the surface area.
Using $n_{\mathrm{eff}}(\mathbf{x})$ for each cell, we define
$L_{\mathrm{cell}}(\mathbf{x})
=H\sqrt{S_{\mathrm{cell}}(\mathbf{x})/
S_{\mathrm{ref}}(n_{\mathrm{eff}})}$.
Here, $H$ is the tree height, and
$S_{\mathrm{ref}}(n_{\mathrm{eff}})$ is the corresponding total
surface area obtained by interpolating the reference geometries for
integer branching orders $n=1,\dots,8$.

\subsection{Cell-wise evaluation of $C_D$}

Once $n_{\mathrm{eff}}$ and $Re_{\mathrm{eff}}$ are determined, the
cell-wise drag coefficient is obtained by evaluating the a priori
relationship at these effective quantities:
\[
C_D(\mathbf{x})
=
C_D\!\left(
n_{\mathrm{eff}}(\mathbf{x}),
Re_{\mathrm{eff}}(\mathbf{x})
\right).
\]
Although the complete reference-tree geometries are defined at
discrete branching orders, the lookup table is represented over a
continuous range of $n$ to allow interpolation using the inferred
non-integer values of $n_{\mathrm{eff}}$. For cells where the inferred
value falls below the single-cylinder limit
($n_{\mathrm{eff}}<1$), we evaluate $C_D$ by interpolating between
the value at $n=1$ and $C_D=0$ at $n_{\mathrm{min}}$, where
$n_{\mathrm{min}}$ corresponds to the empty-cell limit
($\Omega=0$).

\section{Numerical method and simulation setup}
\label{sec:simulation setup}

\subsection{Simulation procedure}

\begin{figure}[htbp]
\centering
\resizebox{0.92\columnwidth}{!}{%
\begin{tikzpicture}[ >=Latex, node distance=5mm, box/.style={ draw, rounded corners=2pt, align=center, minimum width=38mm, minimum height=8mm, font=\footnotesize, inner sep=3pt }, decision/.style={ draw, diamond, aspect=2.2, align=center, font=\footnotesize, inner sep=1.5pt }, io/.style={ draw, trapezium, trapezium left angle=75, trapezium right angle=105, align=center, minimum width=38mm, minimum height=8mm, font=\footnotesize, inner sep=3pt }, stage/.style={ draw, dashed, rounded corners=3pt, inner sep=5pt }, line/.style={ ->, line width=0.55pt } ]

\node[io] (geom)
{Input: Fractal tree geometry};

\node[box, below=of geom] (porous)
{Construct porous-media representation};

\node[box, below=of porous] (geometry)
{Evaluate geometry-dependent quantities\\
$\mathrm{PAD}_i(\mathbf{x}),\quad n_{\mathrm{eff}}(\mathbf{x}),\quad L_{\mathrm{cell}}(\mathbf{x})$};

\node[stage,
      fit=(porous)(geometry),
      label={[font=\footnotesize]right:Preprocessing}]
      (preprocess) {};

\node[box, below=10mm of geometry] (init)
{Initialize velocity field};

\node[box, below=of init] (re)
{Evaluate
$Re_{\mathrm{eff}}(\mathbf{x})$};

\node[box, below=of re] (cd)
{Evaluate
$C_D(\mathbf{x})$};

\node[box, below=of cd] (source)
{Evaluate porous-tree source terms};

\node[box, below=of source] (rans)
{Solve RANS equations};

\node[decision, below=7mm of rans] (conv)
{Converged?};

\node[io, below=7mm of conv] (out)
{Output: Final flow field};

\draw[line] (geom) -- (porous);
\draw[line] (porous) -- (geometry);
\draw[line] (geometry) -- (init);
\draw[line] (init) -- (re);
\draw[line] (re) -- (cd);
\draw[line] (cd) -- (source);
\draw[line] (source) -- (rans);
\draw[line] (rans) -- (conv);
\draw[line] (conv) -- node[right,font=\footnotesize]{yes} (out);

\draw[line]
(conv.west) -- node[above,pos=0.22,font=\footnotesize]{no} ++(-20mm,0)
|- (re.west);

\node[stage,
      fit=(init)(re)(cd)(source)(rans)(conv),
      label={[font=\footnotesize,xshift=5mm]right:RANS iteration}]
      (iteration) {};
\path ([xshift=-14mm]current bounding box.west)
      -- (current bounding box.west);

\end{tikzpicture}%
}
\caption{\label{fig:simulation_procedure}
Simulation workflow for the proposed variable-drag model.
Geometry-dependent quantities are evaluated during preprocessing,
whereas $Re_{\mathrm{eff}}$ and $C_D$ are updated with the local
velocity during the RANS iteration.}
\end{figure}
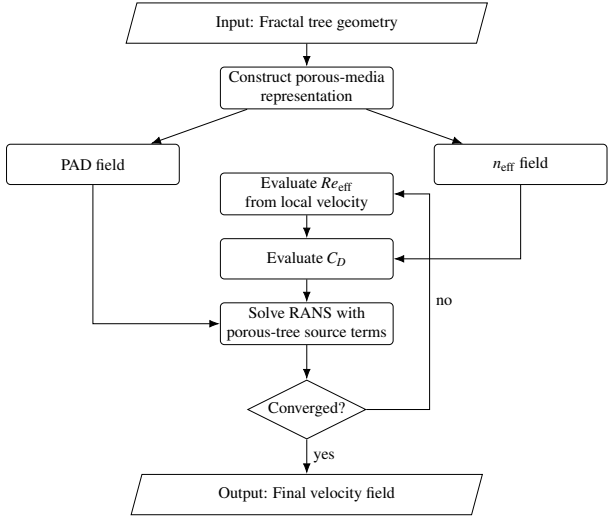

We evaluate the proposed variable-drag model using steady,
incompressible, isothermal RANS simulations for the target tree
geometry (Tree~A). Figure~\ref{fig:simulation_procedure} summarizes
the simulation workflow. Prior to the RANS solution, we construct the
porous-media representation and evaluate the geometry-dependent
quantities, including $\mathrm{PAD}_i(\mathbf{x})$, $n_{\mathrm{eff}}(\mathbf{x})$, and
$L_{\mathrm{cell}}(\mathbf{x})$, as described in Sec.~\ref{sec:drag_model}.
After initializing the velocity field, $Re_{\mathrm{eff}}(\mathbf{x})$ is evaluated
and the corresponding $C_D(\mathbf{x})$ is obtained from the a priori
relationship. The porous-tree source terms are then evaluated and the
RANS equations are solved. Because $Re_{\mathrm{eff}}(\mathbf{x})$ depends on the
local velocity, $Re_{\mathrm{eff}}(\mathbf{x})$ and $C_D(\mathbf{x})$ are updated during the
iterative solution until convergence.

\subsection{Governing equations and porous-tree source terms}

We solve the steady, incompressible, isothermal Reynolds-averaged Navier--Stokes equations with the standard $k$-$\varepsilon$ turbulence model. The transport equations of mass, mean momentum, turbulent kinetic energy (TKE), and its dissipation rate are given as
\begin{equation}\label{eq:mass-transport}
\frac{\partial u_i}{\partial x_i}=0,
\end{equation}

\begin{equation}\label{eq:mom-transport}
\begin{split}
u_j\frac{\partial u_i}{\partial x_j}
&=
-\frac{1}{\rho}\frac{\partial}{\partial x_i}
\left(p+\frac{2}{3}\rho k\right) \\
&\quad+
\frac{\partial}{\partial x_j}
\left[
(\nu+\nu_t)
\left(
\frac{\partial u_i}{\partial x_j}
+
\frac{\partial u_j}{\partial x_i}
\right)
\right]
+S_{u,i},
\end{split}
\end{equation}

\begin{equation}\label{eq:k-transport}
u_j\frac{\partial k}{\partial x_j}=
\frac{\partial}{\partial x_j}
\left[
\left(\nu+\frac{\nu_t}{\sigma_k}\right)
\frac{\partial k}{\partial x_j}
\right]
+
P_k-\varepsilon+S_k,
\end{equation}

\begin{equation}\label{eq:eps-transport}
u_j\frac{\partial \varepsilon}{\partial x_j}=
\frac{\partial}{\partial x_j}
\left[
\left(\nu+\frac{\nu_t}{\sigma_\varepsilon}\right)
\frac{\partial \varepsilon}{\partial x_j}
\right]
+
C_{\varepsilon1}P_k\frac{\varepsilon}{k}
-
C_{\varepsilon2}\frac{\varepsilon^2}{k}
+S_\varepsilon.
\end{equation}
Here, $u_i$ $[\mathrm{m~s^{-1}}]$ is the mean velocity in the $i$th direction, $p$ $[\mathrm{Pa}]$ is the mean pressure, $k$ $[\mathrm{m^2~s^{-2}}]$ is the TKE, and $\varepsilon$ $[\mathrm{m^2~s^{-3}}]$ is the TKE dissipation rate. The turbulent viscosity in the standard $k$-$\varepsilon$ model is $\nu_t=C_\mu k^2/\varepsilon$, and $P_k=\nu_t\left(\frac{\partial u_i}{\partial x_j}+\frac{\partial u_j}{\partial x_i}\right)\frac{\partial u_i}{\partial x_j}$
denotes the TKE production term. The remaining variables and coefficients are constants in this study: air density $\rho$ $[\mathrm{kg~m^{-3}}]$, kinematic viscosity of air $\nu$ $[\mathrm{m^2~s^{-1}}]$, and turbulence-model coefficients ($C_\mu=0.09$, $C_{\varepsilon1}=1.44$, $C_{\varepsilon2}=1.92$, $\sigma_k=1$, $\sigma_\varepsilon=1.3$).

The final terms on the right-hand side of Eqs.~\eqref{eq:mom-transport}-\eqref{eq:eps-transport} represent the porous-tree source terms. Following Endalew et al.~\citep{Endalew2009IJHFF} and Oshio et al.~\citep{Oshio2021}, we introduce them as
\begin{equation}\label{eq:Su}
S_{u,i}=-\frac{1}{2}C_D\,\mathrm{PAD}_i\,|\mathbf{u}|u_i,
\end{equation}

\begin{equation}\label{eq:Sk}
S_k=\frac{1}{2}C_D\,\mathrm{PAD}_{\mathrm{ave}}\,|\mathbf{u}|k,
\end{equation}

\begin{equation}\label{eq:Seps}
S_\varepsilon=\frac{1}{2}C_D\,\mathrm{PAD}_{\mathrm{ave}}\,|\mathbf{u}|\varepsilon.
\end{equation}
Here, $|\mathbf{u}|~[\mathrm{m\,s^{-1}}]$ is the magnitude of the velocity vector at each cell, 
$\mathrm{PAD}_i~[\mathrm{m^{-1}}]$ is the plant area density in the $i$th direction, 
$\mathrm{PAD}_{\mathrm{ave}}~[\mathrm{m^{-1}}]$ is the velocity-weighted average plant area density, 
and $C_D$ is the drag coefficient. In the constant-$C_D$ cases, we prescribe $C_D$ as a uniform constant, whereas in the proposed model we evaluate it cell-wise.

\subsection{Computational domain, boundary conditions, and discretization}

\begin{figure}[htbp]
\centering
\includegraphics[width=\linewidth]{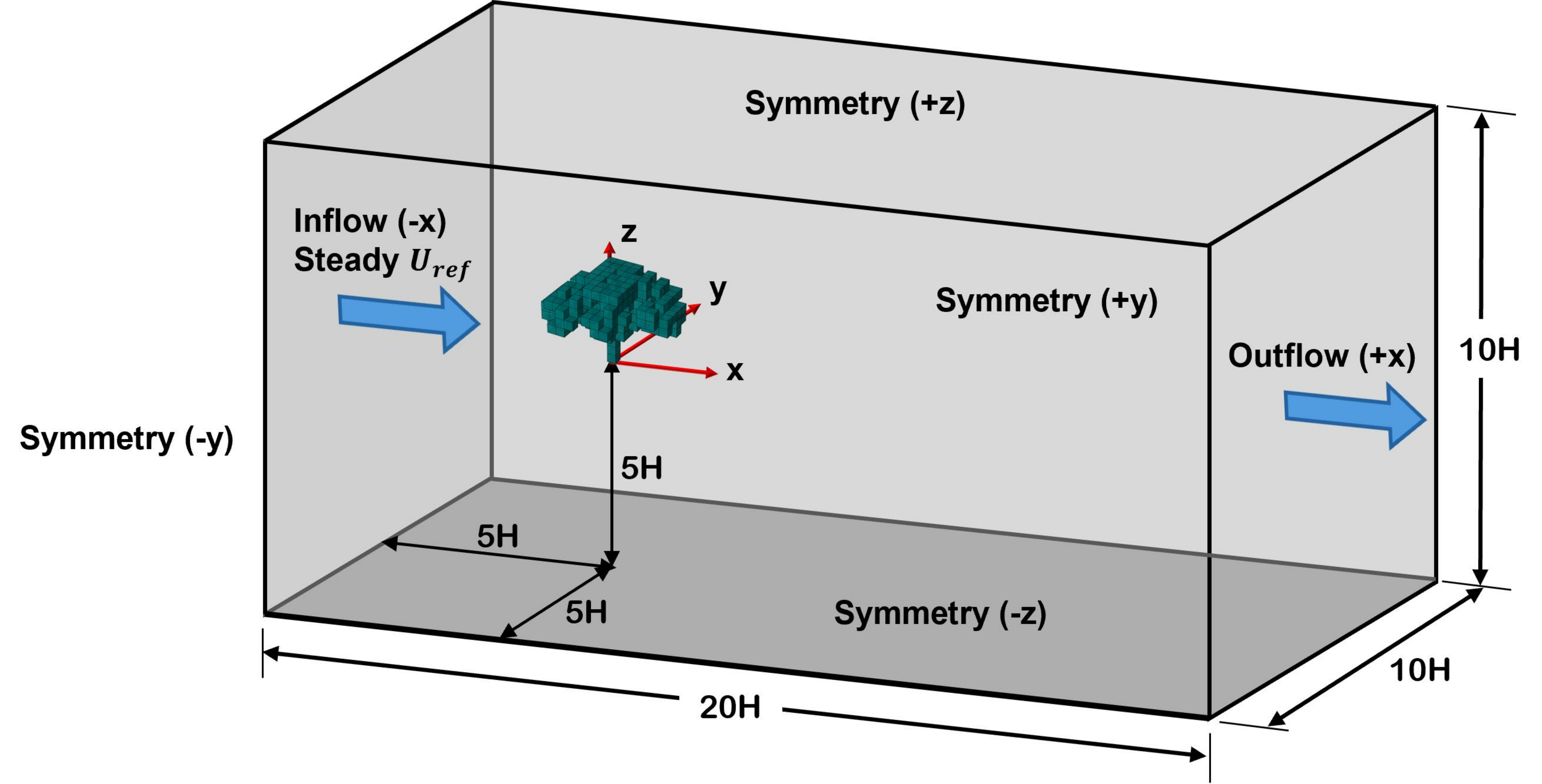}
\caption{\label{fig:domain_setup}
Computational domain and boundary-condition configuration.}
\end{figure}

We place Tree~A in a rectangular computational domain of size $20H\times10H\times10H$ in the $(x,y,z)$ directions, where $x$ is the inflow direction. We locate the tree at a distance of $5H$ from the inflow plane and at the center of the $y$--$z$ plane, i.e., at equal distances from the $\pm y$ and $\pm z$ boundaries. The lateral and top and bottom boundaries are therefore sufficiently far from the tree that wall-confinement effects are not expected to influence the flow around the tree~\citep{Tominaga2008, Franke2007}. Figure~\ref{fig:domain_setup} presents the computational domain and boundary-condition configuration.

We prescribe inflow and outflow boundaries in the $x$ direction, while we treat the lateral and top and bottom boundaries as symmetry boundaries. In this study, we normalize the tree height as $H=1\,\mathrm{m}$. We consider six grid resolutions, $\Delta/H = 1,\;1/2,\;1/4,\;1/6,\;1/8,\;1/10$.

We perform the simulations using OpenFOAM v11~\citep{openfoam_v11} with the steady-state solver \texttt{simpleFoam} and a SIMPLE-type pressure-velocity coupling procedure. We use predominantly second-order spatial discretization: central-type schemes for gradients and diffusion terms, and upwind-biased schemes for convection terms, with bounded treatment for $k$ and $\varepsilon$. We regard the solution as converged when the residuals fall below $10^{-4}$ for velocity, $k$, and $\varepsilon$, and $10^{-3}$ for pressure $p$.

\begin{figure}[htbp]
\centering

\begin{subfigure}[t]{0.3\linewidth}
  \centering
  \includegraphics[width=\linewidth]{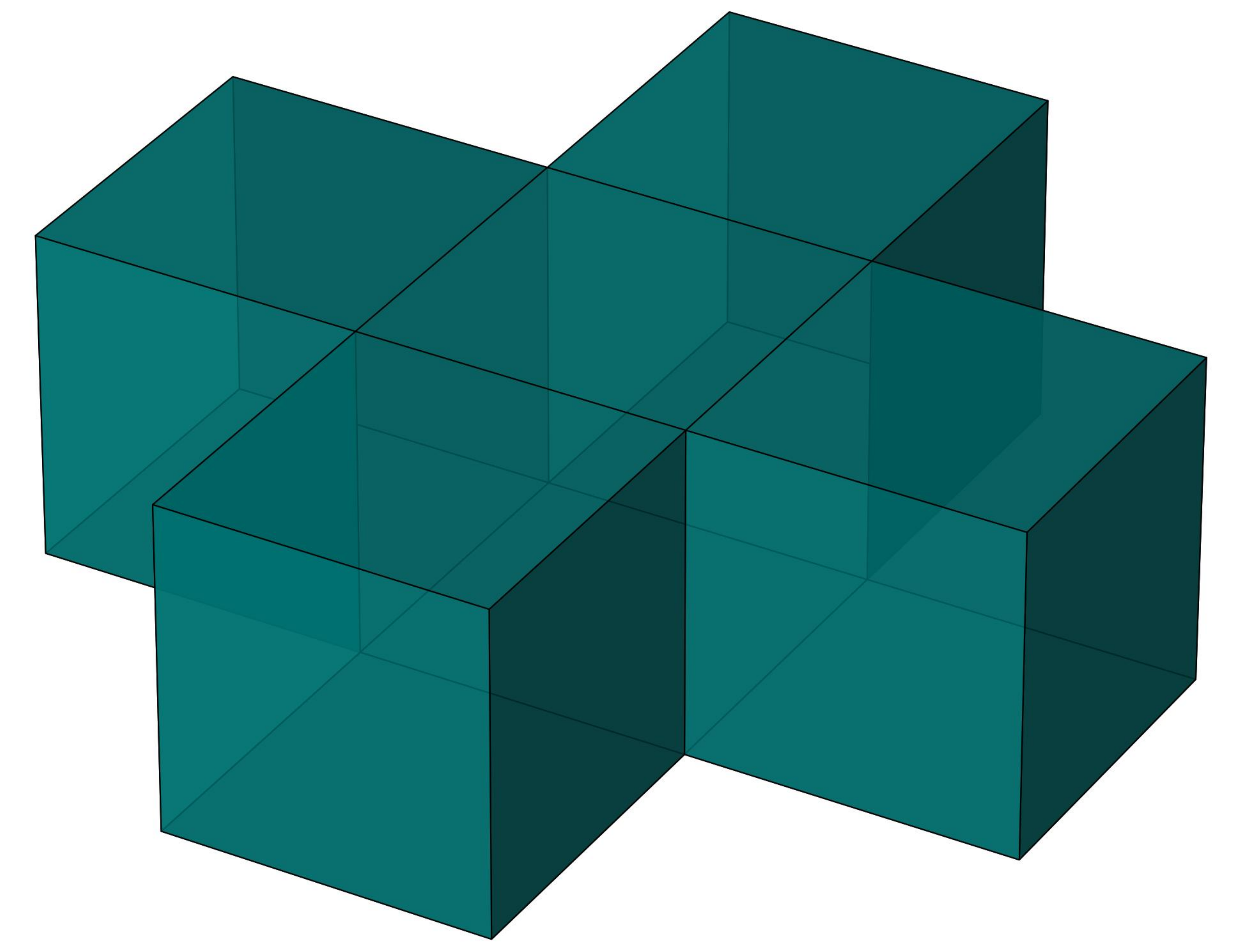}
  \caption{$H/1$}
\end{subfigure}\hspace{0.03\linewidth}
\begin{subfigure}[t]{0.3\linewidth}
  \centering
  \includegraphics[width=\linewidth]{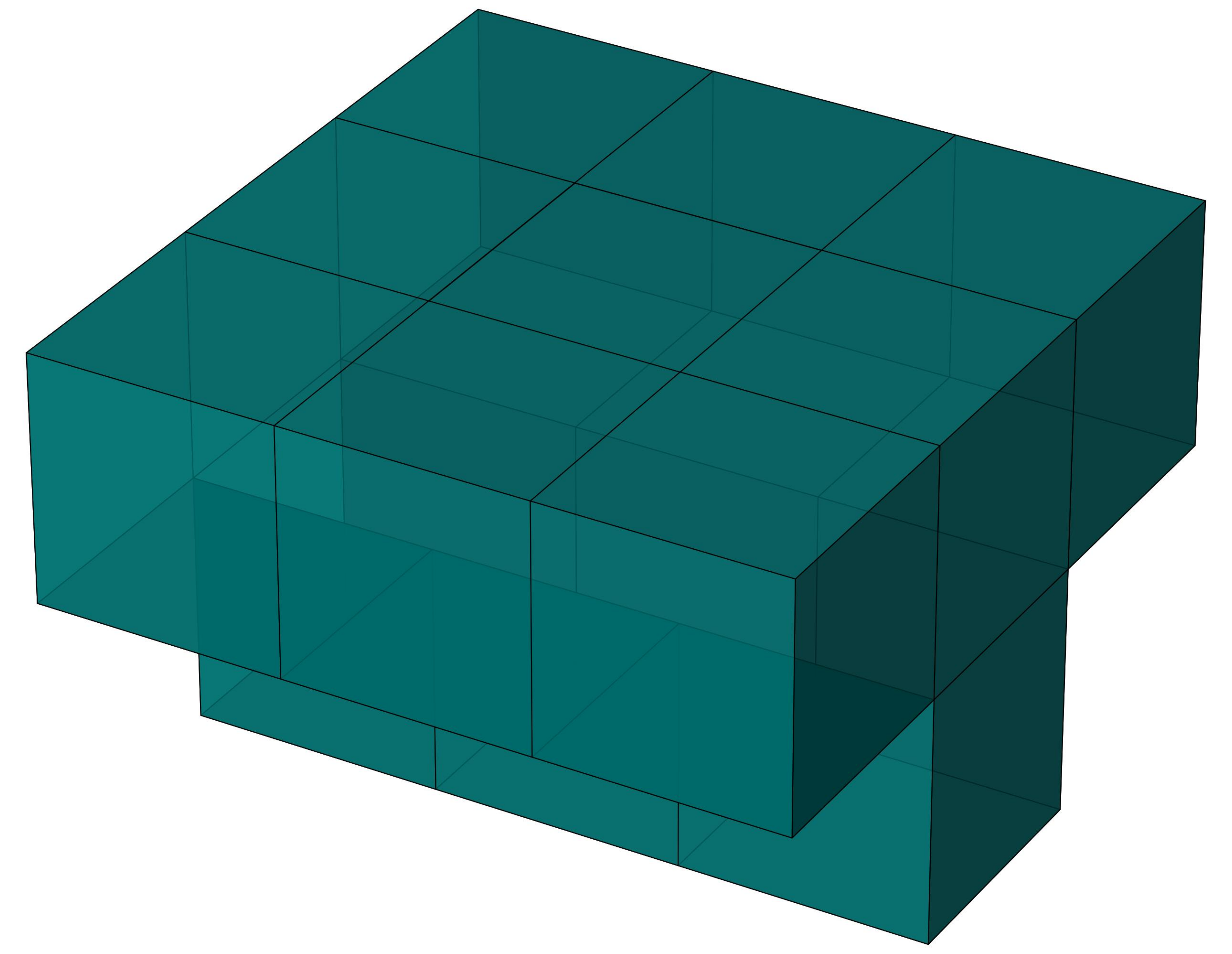}
  \caption{$H/2$}
\end{subfigure}
\begin{subfigure}[t]{0.3\linewidth}
  \centering
  \includegraphics[width=\linewidth]{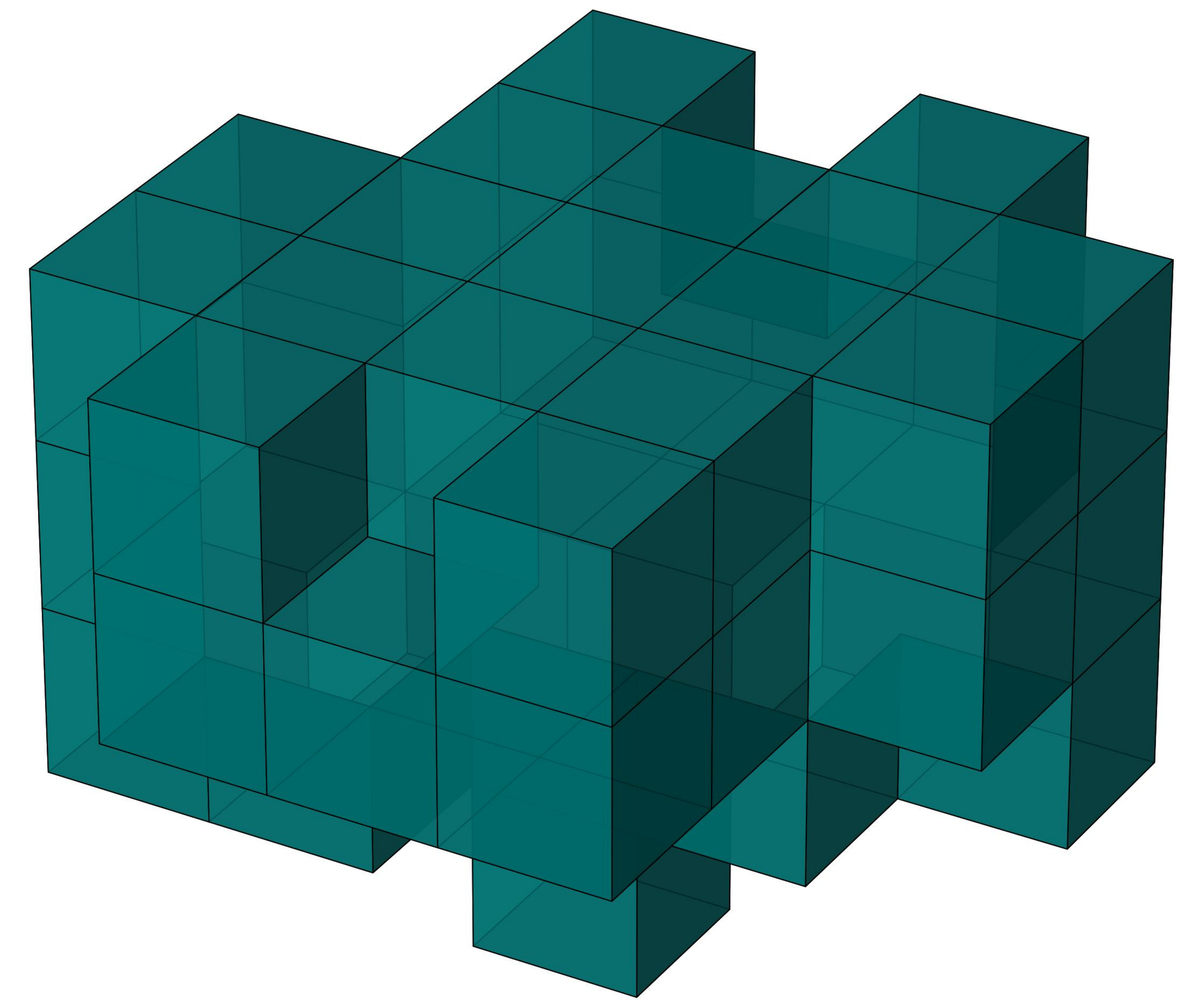}
  \caption{$H/4$}
\end{subfigure}

\vspace{0.8ex}

\begin{subfigure}[t]{0.3\linewidth}
  \centering
  \includegraphics[width=\linewidth]{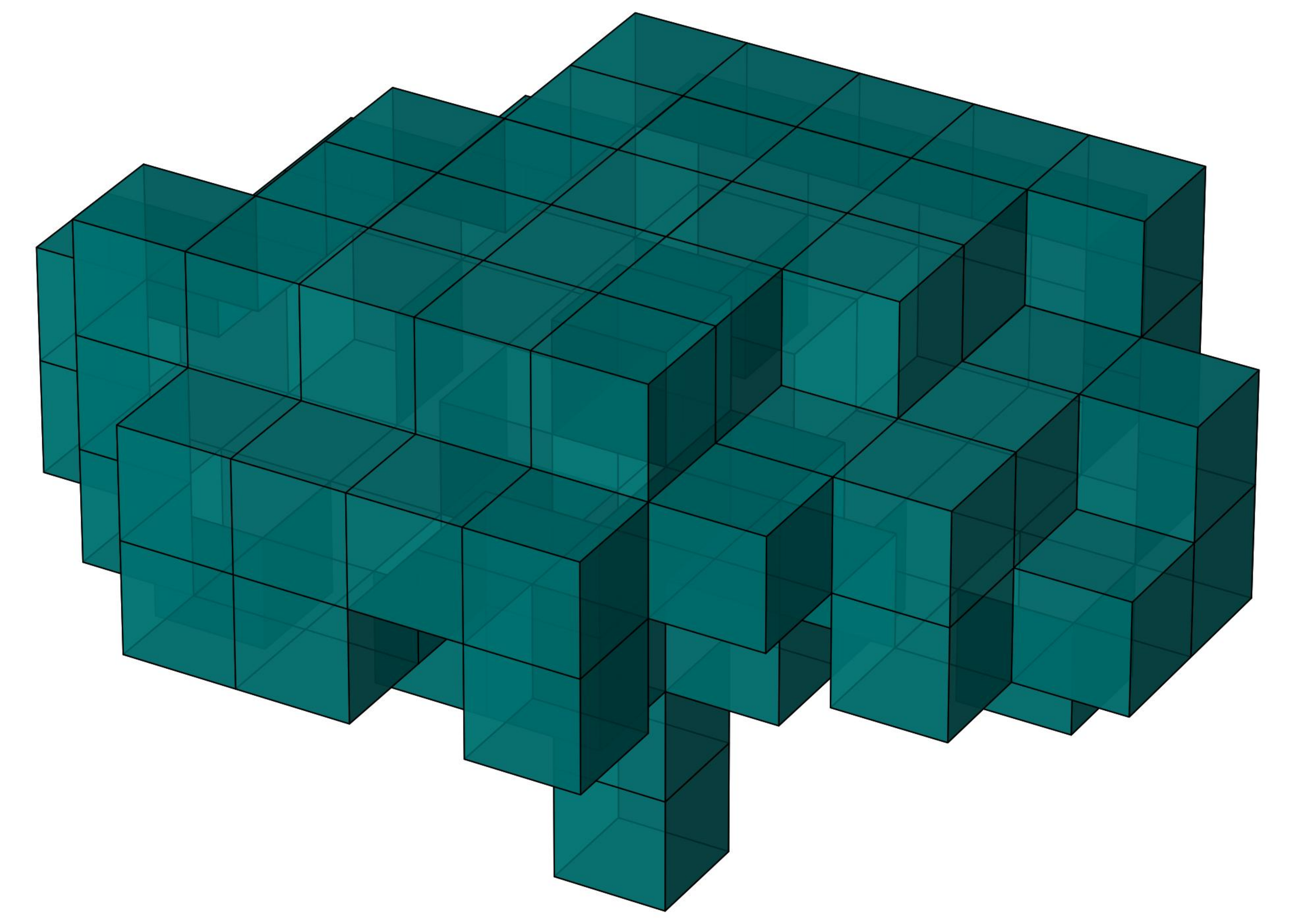}
  \caption{$H/6$}
\end{subfigure}
\begin{subfigure}[t]{0.3\linewidth}
  \centering
  \includegraphics[width=\linewidth]{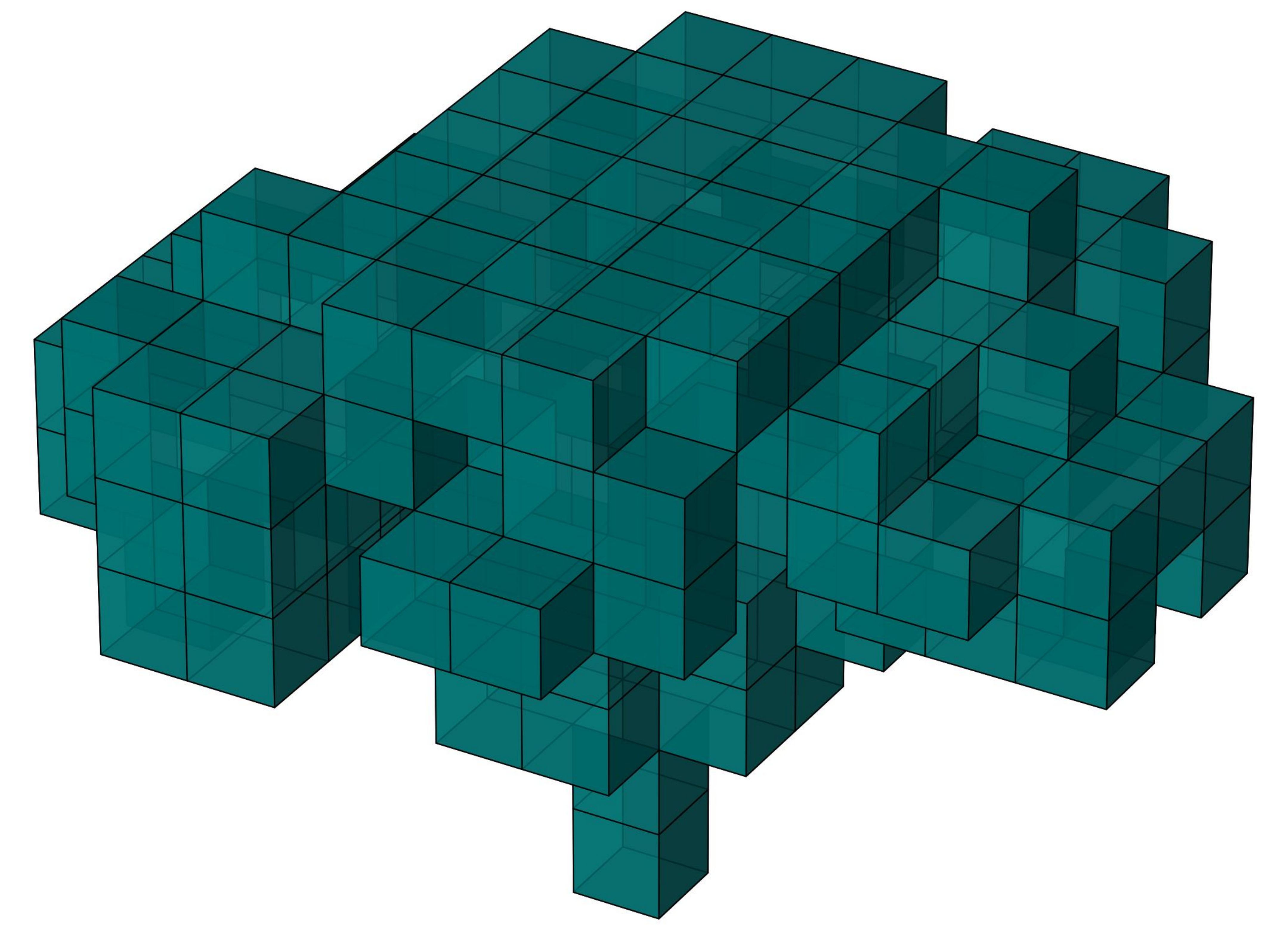}
  \caption{$H/8$}
\end{subfigure}\hspace{0.03\linewidth}
\begin{subfigure}[t]{0.3\linewidth}
  \centering
  \includegraphics[width=\linewidth]{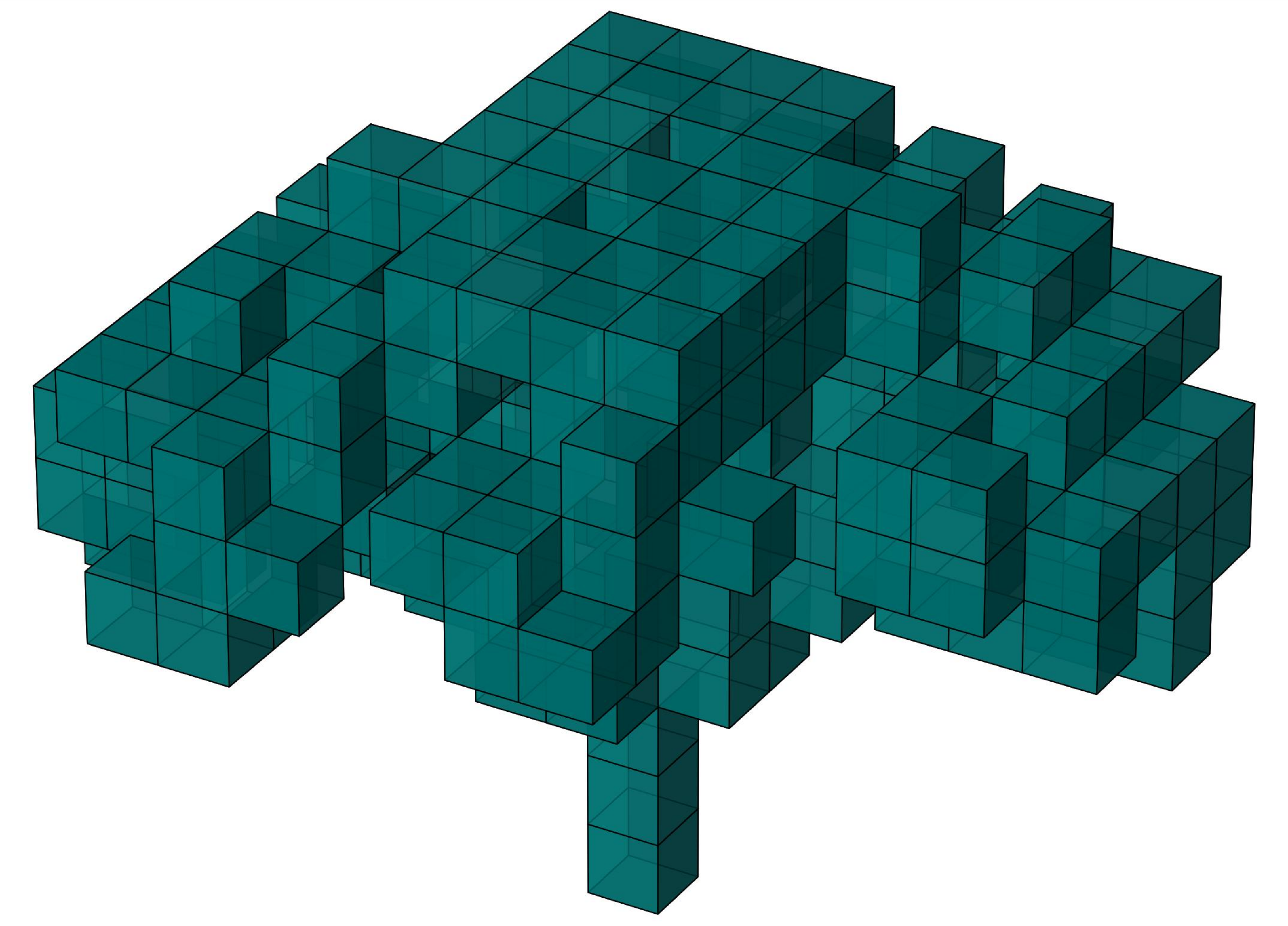}
  \caption{$H/10$}
\end{subfigure}
\caption{\label{fig:tree_representation}
Tree-mask representation of Tree~A at each grid resolution.}
\end{figure}

To illustrate how the porous-tree region is represented on the computational grid, we present the tree mask of Tree~A at each grid resolution in Fig.~\ref{fig:tree_representation}.

\subsection{Simulation cases and evaluation metric}

We consider three global Reynolds numbers based on tree height: $Re_H \approx 2{,}600$, $6{,}600$, and $66{,}000$.
We compare three model classes: (i) a general conventional model using uniform PAD with constant $C_D$, (ii) an advanced conventional model using voxel-resolved PAD with constant $C_D$, and (iii) the proposed model using voxel-resolved PAD with $C_D(n_{\mathrm{eff}},Re_{\mathrm{eff}})$. For the constant-$C_D$ models, we selected $C_D = 0.5$, $1.0$, and $1.5$ based on ranges reported in earlier works~\citep{Katul2004}. Table~\ref{tab:simulation_matrix} summarizes the simulation matrix.

We evaluate model performance primarily using aerodynamic porosity, defined as the ratio of leeward to windward plane-averaged streamwise velocity across the tree. We adopt this quantity because the present study focuses on the reproduction of bulk drag rather than on detailed wake structure. Although wake-velocity distributions are also informative, they are inevitably sensitive to grid resolution through advection and numerical diffusion, which makes it difficult to isolate the effect of the drag model itself. Aerodynamic porosity therefore provides a more suitable metric for the present purpose, while a more detailed assessment of wake structure remains an important subject for future work.

We compute aerodynamic porosity using plane-averaged streamwise velocity on two planes located at the upstream and downstream edges of the tree, i.e., at $x/H=-0.6$ and $x/H=+0.6$, respectively. For the grid-robustness assessment, we use a $1.5H\times1.0H$ averaging plane, which is the minimum window that contains the projected area of the porous tree representation for all grid resolutions considered in this study. For the Reynolds-number-dependence assessment against the DNS reference, we use a $1.024H\times1.0H$ averaging plane so that the AP definition is consistent with the reference study of the exact tree geometry~\citep{Yin2025}.

\begin{table}[htbp]
\centering
\caption{\label{tab:simulation_matrix}
Simulation matrix used in the present RANS study. All model classes are evaluated for $H/\Delta=1,\,2,\,4,\,6,\,8,\,10$ and $Re_H = 2{,}600$, $6{,}600$, $66{,}000$.}
\begin{tabular}{lll}
\hline
Model class & PAD field & $C_D$ model \\
\hline
General conventional & Uniform & Constant ($0.5,\,1.0,\,1.5$) \\
Advanced conventional & Voxel-resolved & Constant ($0.5,\,1.0,\,1.5$) \\
Proposed & Voxel-resolved & $C_D(n_{\mathrm{eff}},Re_{\mathrm{eff}})$ \\
\hline
\end{tabular}
\end{table}

\section{Results and discussion}
\label{sec:RaD}

\subsection{Qualitative verification of the simulated flow field}

\begin{figure}[htbp]
\centering

\begin{subfigure}[t]{\linewidth}
  \centering
  \includegraphics[width=\linewidth]{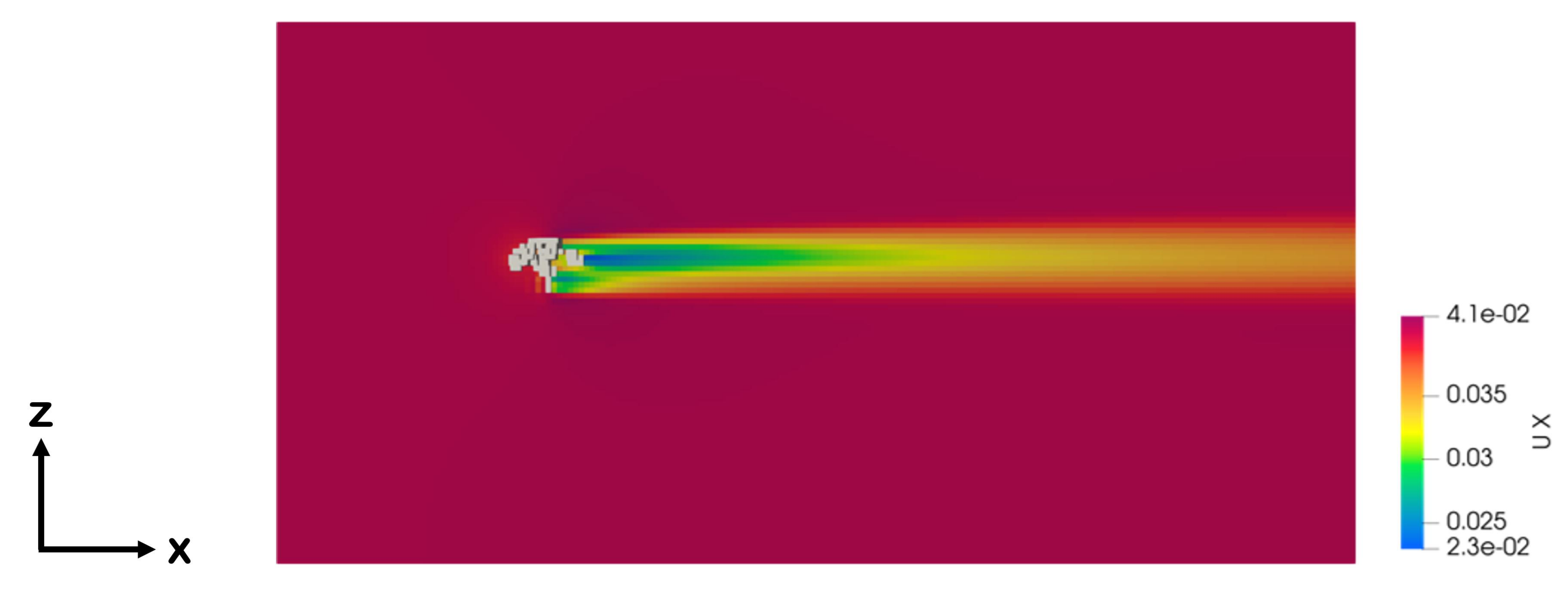}
  \caption{Vertical center plane ($y=0$)}
  \label{fig:vel_y0}
\end{subfigure}

\vspace{0.8ex}

\begin{subfigure}[t]{\linewidth}
  \centering
  \includegraphics[width=\linewidth]{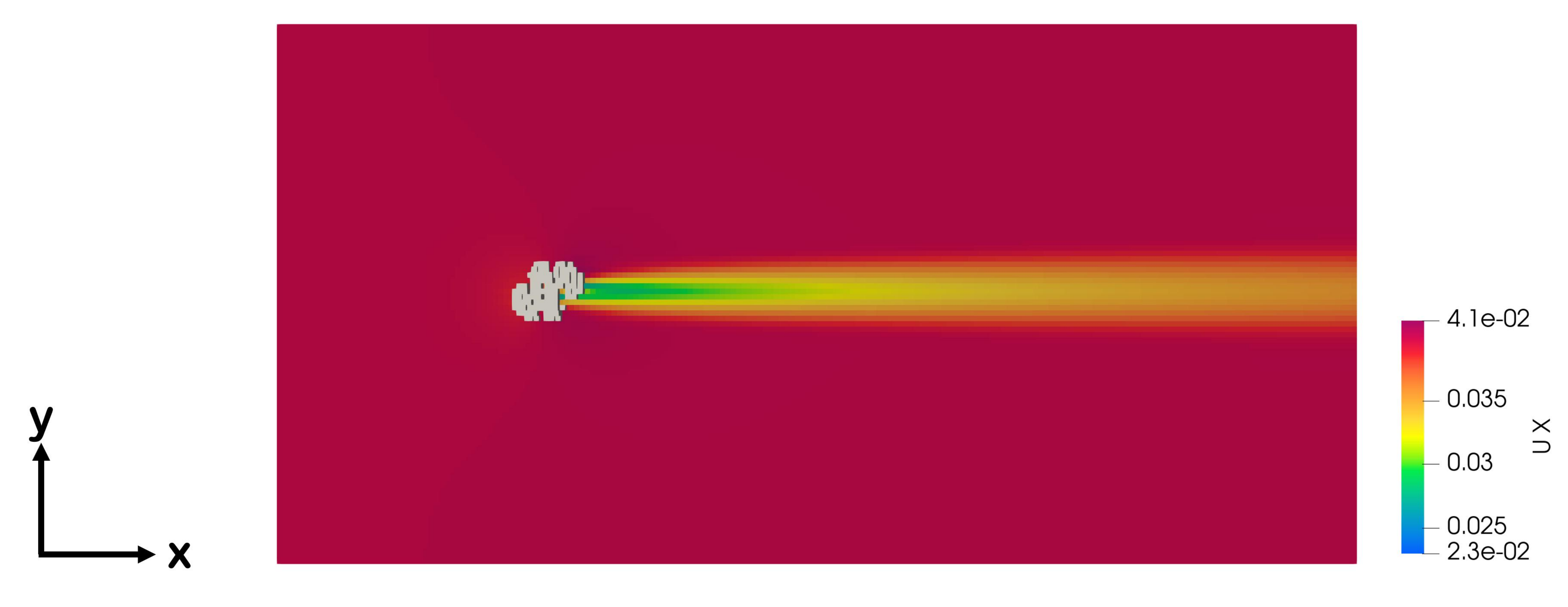}
  \caption{Horizontal mid-height plane ($z=0.5H$)}
  \label{fig:vel_z05H}
\end{subfigure}

\caption{\label{fig:velocity_contours}
Contours of streamwise velocity for the proposed model at $Re_H=2{,}600$ and $H/\Delta=10$.
}
\end{figure}

Figure~\ref{fig:velocity_contours} presents contours of the streamwise velocity for the proposed model at $Re_H=2{,}600$ and $H/\Delta=10$, shown in the vertical center plane ($y=0$) and the horizontal mid-height plane ($z=0.5H$).

In the vertical center plane, the flow exhibits a clear velocity deficit within and behind the porous tree region, followed by gradual wake recovery in the downstream direction. The flow also accelerates around and above the tree, indicating momentum redistribution around a resistive canopy-like obstacle. In the horizontal mid-height plane, a clear velocity deficit is similarly observed within the porous tree region, extending into a relatively narrow wake downstream. These features are qualitatively consistent with the fundamental aerodynamic behavior expected for a tree represented as a momentum-extracting porous body~\citep{gromke2008aerodynamic, sanz2003note, mochida2008examining}. This qualitative verification confirms that the proposed model produces a plausible flow field while retaining the practical porous-media representation, supporting its use as a physically grounded alternative to ad hoc constant-$C_D$ prescriptions.

\subsection{Grid sensitivity}
In the following quantitative comparisons, we use $(1-\mathrm{AP})$ rather than $\mathrm{AP}$ itself, where $\mathrm{AP}$ denotes aerodynamic porosity. Since $\mathrm{AP}$ is generally close to unity in the present cases, expressing the drag effect as $(1-\mathrm{AP})$ makes the differences among models clearer and more interpretable.

\begin{figure}[htbp]
\centering

\begin{subfigure}[t]{0.7\linewidth}
  \centering
  \includegraphics[width=\linewidth]{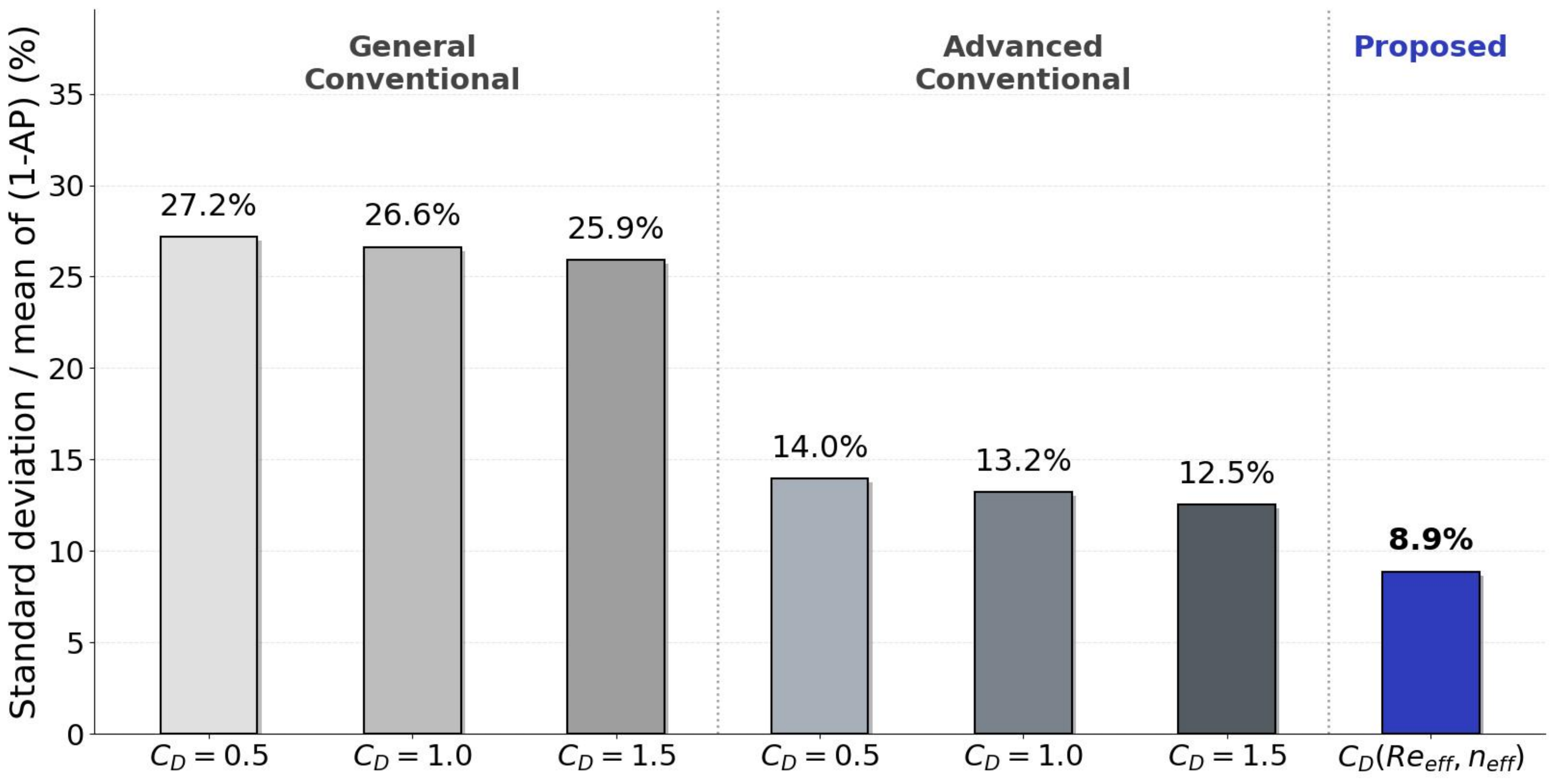}
  \caption{$Re_H=2{,}600$}
  \label{fig:grid_std_u004}
\end{subfigure}

\begin{subfigure}[t]{0.7\linewidth}
  \centering
  \includegraphics[width=\linewidth]{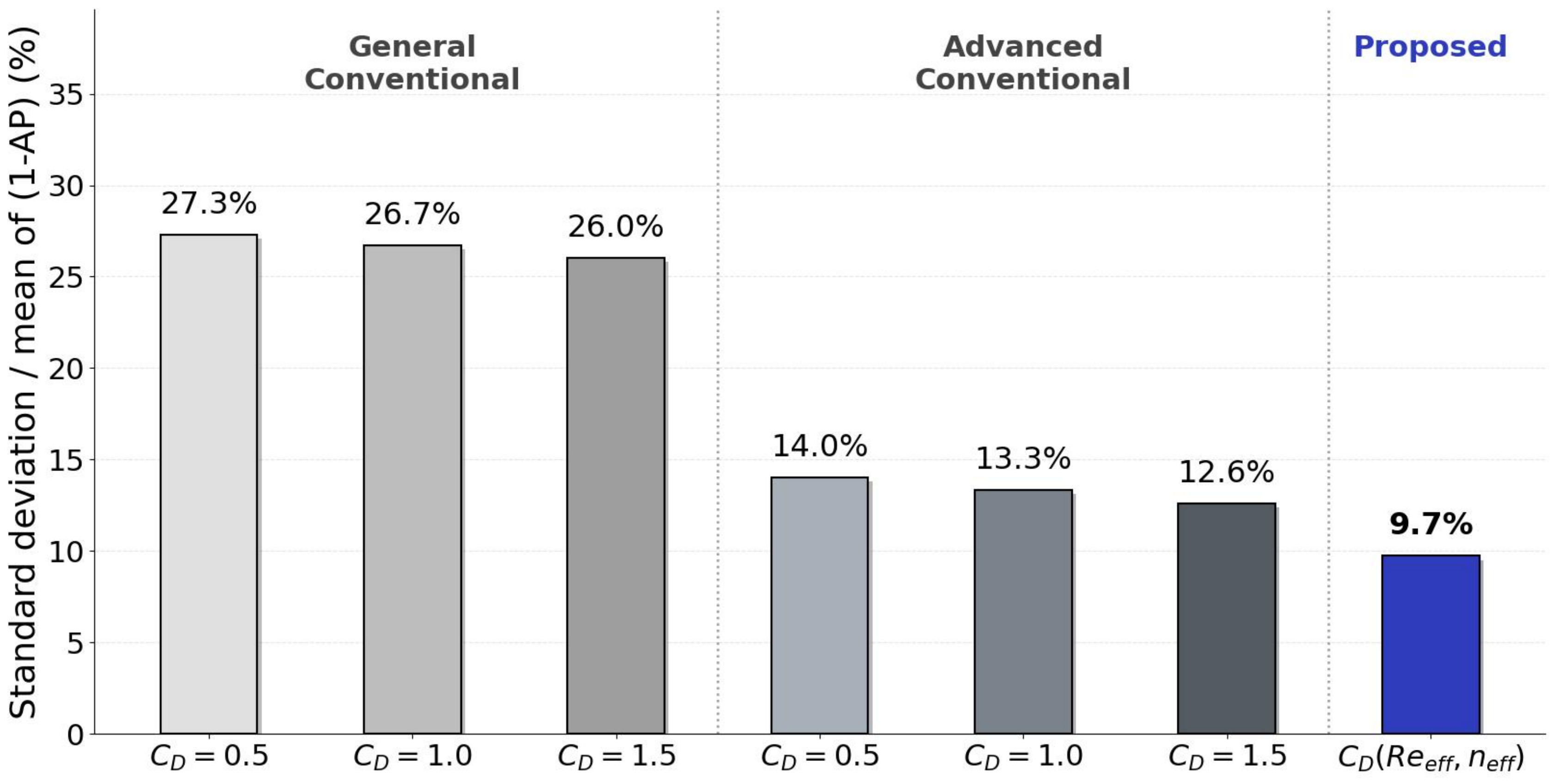}
  \caption{$Re_H=6{,}600$}
  \label{fig:grid_std_u01}
\end{subfigure}

\begin{subfigure}[t]{0.7\linewidth}
  \centering
  \includegraphics[width=\linewidth]{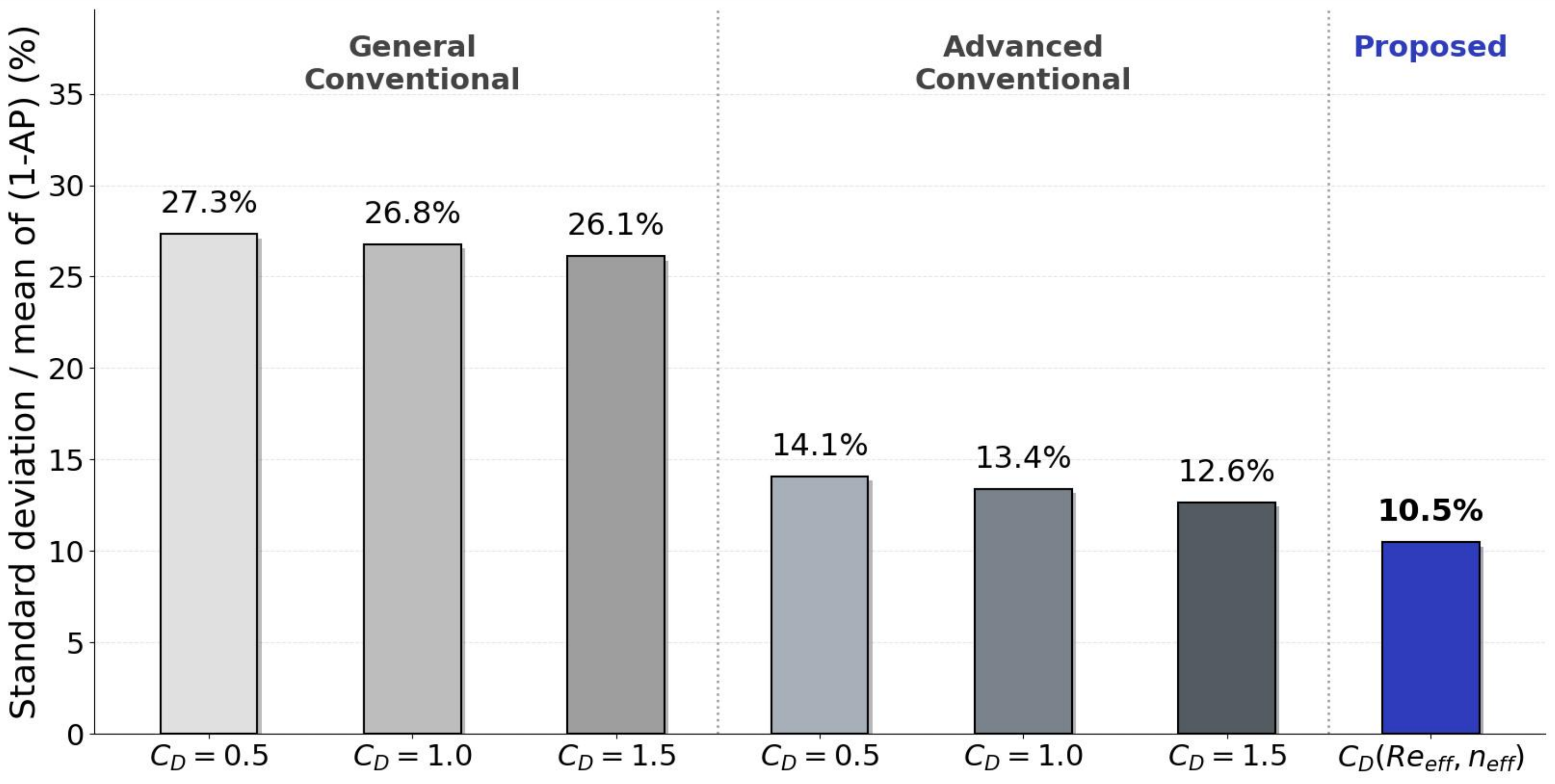}
  \caption{$Re_H=66{,}000$}
  \label{fig:grid_std_u10}
\end{subfigure}

\caption{\label{fig:grid_std}
Standard deviation of $(1-\mathrm{AP})$ across grid resolutions for each model. (a)--(c) correspond to the three inflow cases.}
\end{figure}

\begin{figure}[htbp]
\centering
\includegraphics[width=0.90\linewidth]{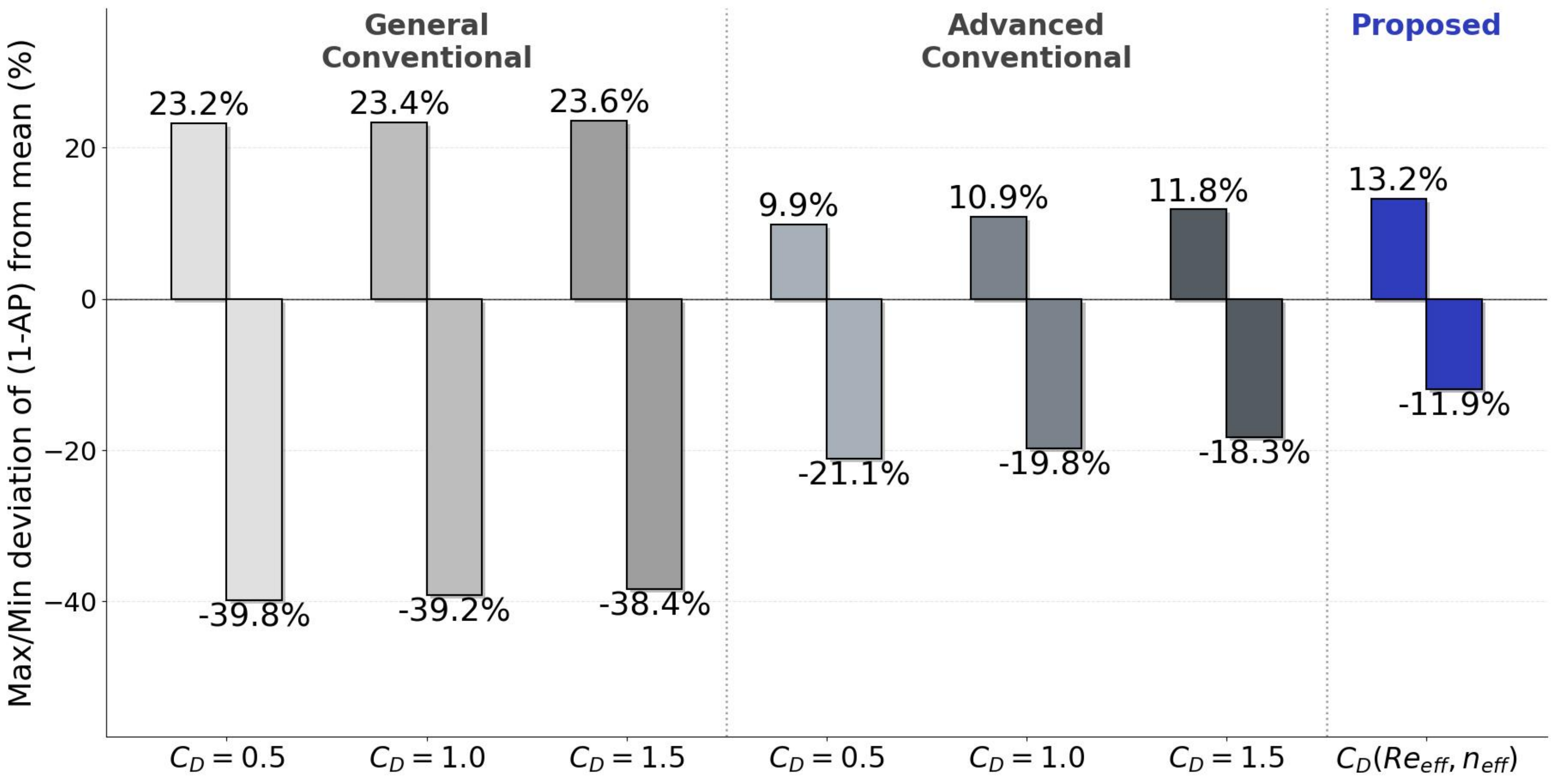}
\caption{\label{fig:grid_minmax}
Minimum and maximum values of $(1-\mathrm{AP})$ across grid resolutions for each model, averaged over the three inflow cases.}
\end{figure}

We first examine how strongly the predicted drag effect depends on grid resolution. Figure~\ref{fig:grid_std} presents, for each model and inflow velocity, the standard deviation of $(1-\mathrm{AP})$ across the six grid resolutions. A smaller value indicates that the predicted bulk drag effect is less sensitive to grid resolution. Figure~\ref{fig:grid_minmax} complements this comparison by presenting the minimum and maximum deviations of $(1-\mathrm{AP})$ across grid resolutions, averaged over the three inflow cases. This metric highlights the spread of the predicted drag effect associated with grid coarsening or refinement.

Figure~\ref{fig:grid_std} shows that the general conventional model yields a standard deviation of approximately $27\%$ for all constant-$C_D$ settings, whereas the advanced conventional model reduces this value to approximately $13\%$. In contrast, the proposed model consistently achieves a smaller deviation of approximately $9\%$ in all inflow cases. Relative to the general conventional model, this corresponds to an improvement in resolution robustness of about $65\%$, while the improvement relative to the advanced conventional model is about $30\%$. Here, these percentages denote relative improvement rates of the deviation metric.

A similar tendency appears in the minimum--maximum range shown in Fig.~\ref{fig:grid_minmax}. The proposed model exhibits a substantially narrower deviation range of $(1-\mathrm{AP})$ across grid resolutions than either the general or advanced conventional model. Quantitatively, the range is reduced by about $60\%$ relative to the general conventional model and by about $20\%$ relative to the advanced conventional model, again in terms of relative improvement rates of the deviation metric.

These results indicate that the proposed formulation provides a more robust representation of bulk tree drag against changes in grid resolution. Specifically, the proposed model not only improves upon simplified conventional approaches that assume uniform PAD together with a constant $C_D$, but also yields smaller resolution-induced variations than advanced conventional models that prescribe voxel-resolved PAD while still retaining a constant $C_D$. This result indicates that incorporating the cell-wise morphological descriptor $n_{\mathrm{eff}}$ into the drag coefficient is important for reducing the dependence of bulk drag prediction on tree discretization.

\subsection{Reynolds-number dependence}

\begin{figure}[htbp]
\centering
\includegraphics[width=0.95\linewidth]{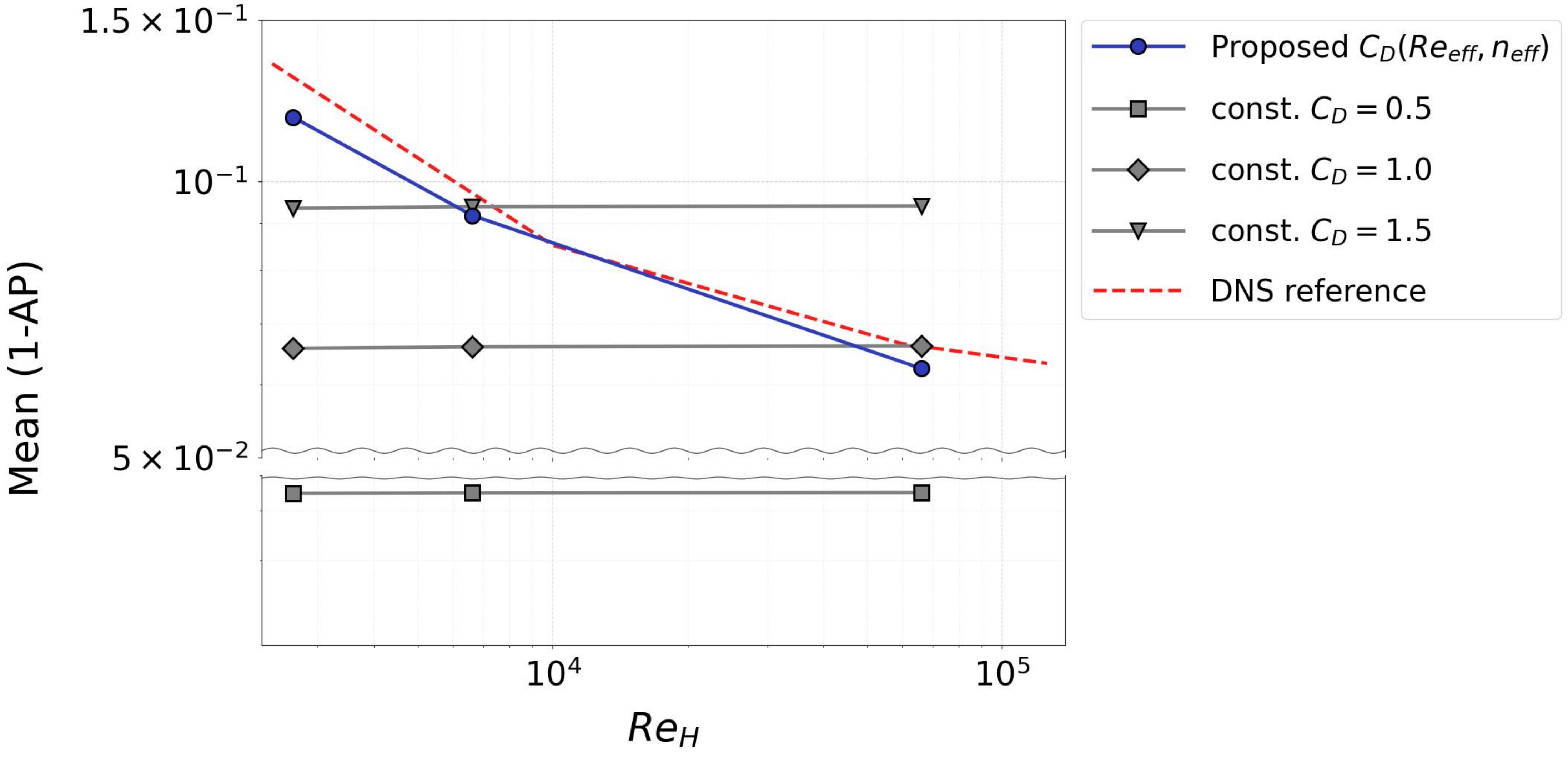}
\caption{\label{fig:re_dependence}
Dependence of $(1-\mathrm{AP})$, averaged over the grid-resolution
cases, on the tree-height-based Reynolds number $Re_H$. The
constant-$C_D$ results correspond to the advanced conventional model,
and the DNS reference data are also shown.}
\end{figure}

Figure~\ref{fig:re_dependence} presents $(1-\mathrm{AP})$, averaged
over the grid-resolution cases, as a function of the
tree-height-based Reynolds number $Re_H$, together with the DNS
reference data~\citep{Yin2025}. For clarity, the advanced conventional
model is shown as the representative constant-$C_D$ comparison. The
constant-$C_D$ model predicts nearly unchanged values of
$(1-\mathrm{AP})$ across the three Reynolds numbers, whereas the DNS
reference exhibits a clear Reynolds-number dependence. In contrast,
the proposed model agrees well with the DNS results in both the
magnitude of $(1-\mathrm{AP})$ and its variation with $Re_H$. This
agreement is obtained without empirical retuning across the inflow
conditions, using the local cell-wise quantities
$n_{\mathrm{eff}}(\mathbf{x})$ and
$Re_{\mathrm{eff}}(\mathbf{x})$.

The a priori $C_D$ table used in the proposed model is constructed
using the analytical drag-estimation method of
\citet{Tokiwa2026}. As reported in that study, the method exhibits a
systematic overprediction of approximately $30$--$40\%$ in the
whole-tree drag coefficient relative to the resolved reference
values. Although this bias is inherited by the table used in the
present RANS simulations, the resulting aerodynamic porosity agrees
well with the DNS results in both magnitude and Reynolds-number
dependence.

A principal reason why the $C_D$ bias is not directly reflected in
AP is the quadratic form of the porous-media momentum sink,
\[
S_{u,i}
=
-\frac{1}{2}C_D\,\mathrm{PAD}_i\,|\mathbf{u}|u_i.
\]
An increase in $C_D$ reduces the local velocity within the porous
region, which partially offsets the increase in the sink magnitude
through the $|\mathbf{u}|u_i$ term. Consequently, the relationship
between the prescribed $C_D$ and AP is nonlinear. The difference in
velocity definitions may also contribute: the whole-tree $C_D$ is
defined using an upstream reference velocity, whereas the
porous-media momentum sink is evaluated using the local velocity in
each cell~\citep{Nepf2012,Pan2016,Zeng2020}. Therefore, an offset in
the whole-tree $C_D$ does not necessarily translate into an
equivalent offset in AP.
Despite the systematic overprediction inherited from the analytical drag-estimation method, the proposed model reproduces the DNS variation of the whole-tree bulk drag effect with Reynolds number across the inflow conditions, without empirical retuning.

\section{Conclusions}
\label{sec:conclusion}

In this study, we proposed and evaluated a fractal-based variable-drag model for porous-media tree representations. By prescribing the drag coefficient cell-wise as $C_D(n_{\mathrm{eff}},Re_{\mathrm{eff}})$, where $n_{\mathrm{eff}}$ and $Re_{\mathrm{eff}}$ denote the cell-effective branching order and the cell-effective Reynolds number, respectively, the model accounts for local morphological complexity and flow regime. The qualitative verification showed that the proposed model produces a plausible aerodynamic response, including a velocity deficit within and behind the porous tree region, bypass flow, and wake recovery. This result supports the proposed formulation as a physically grounded alternative to ad hoc constant-$C_D$ prescriptions conventionally used in porous-tree modeling.

The grid-sensitivity analysis showed that the proposed model improves resolution robustness compared with conventional constant-$C_D$ models. By incorporating the spatial variation of $n_{\mathrm{eff}}$, the model produced more consistent aerodynamic porosity across grid resolutions, indicating that local morphological information reduces the dependence of bulk drag prediction on tree discretization.

The Reynolds-number-dependence analysis showed that the proposed model reproduces the variation of the bulk drag effect across inflow conditions without empirical retuning. Importantly, the global Reynolds-number dependence of the whole-tree aerodynamic response was recovered from local cell-wise quantities, namely $n_{\mathrm{eff}}(\mathbf{x})$ and $Re_{\mathrm{eff}}(\mathbf{x})$. This result demonstrates that the proposed model improves adaptability to varying flow conditions while retaining the computational efficiency of porous-media representations.

Several directions for future work follow naturally from these results. Although the present numerical assessment was restricted to steady RANS simulations, the drag formulation itself is not limited to steady calculations. An important methodological extension is therefore to apply the present model in unsteady CFD, particularly LES, in order to examine its ability to capture time-dependent wake dynamics and turbulence structure. Another important direction is to extend the present model from a single isolated tree to multiple-tree configurations and eventually to urban-scale simulations, which will be essential for assessing its applicability to real urban micrometeorology, including ventilation, thermal environment, and scalar transport.

\section*{CRediT authorship contribution statement}
T. Tokiwa: Conceptualization, Methodology, Software, Validation, Formal analysis, Data curation, Visualization, Writing -- original draft. 
Y. Yin: Methodology, Software, Formal analysis, Writing -- review and editing.
R. Onishi: Conceptualization, Supervision, Resources, Funding acquisition, Writing -- review and editing.

\section*{Declaration of competing interest}
The authors declare that they have no known competing financial interests or
personal relationships that could have appeared to influence the work reported
in this paper.

\section*{Data availability}
The data that support the findings of this study are available from
the corresponding author upon reasonable request.

\appendix
\section{Theoretical prediction of the grid-dependent effective
branching order}
\label{app:neff_prediction}

This appendix derives the theoretical prediction
$n_{\mathrm{eff}}^{\mathrm{th}}$ used as a grid-dependent reference
for comparison with the cell-wise effective branching order inferred
from the $\Omega$--$n$ relationship. 

Let $H$ denote the tree height and define the grid factor as
\begin{equation}
k\equiv\frac{H}{\Delta}.
\label{eq:grid_factor}
\end{equation}
The geometrical space-filling property at scale $\Delta$ is
characterized by box counting. The number of occupied boxes satisfies
\begin{equation}
N(\Delta)
\propto
\left(\frac{H}{\Delta}\right)^{D_{\mathrm{eff}}}
=
k^{D_{\mathrm{eff}}},
\label{eq:boxcount_k}
\end{equation}
where $D_{\mathrm{eff}}$ is the box-counting exponent over the scale
range considered.

Under the present coarse-graining model, an increase of one branching
order corresponds to a scale change by a factor $b>1$. The theoretical
effective branching order at grid scale $\Delta$ is therefore written
as
\begin{equation}
n_{\mathrm{eff}}^{\mathrm{th}}(\Delta)
=
n_0-\log_b N(\Delta),
\label{eq:neff_th_def}
\end{equation}
where $n_0$ is the branching order of the complete tree. Substituting
Eq.~\eqref{eq:boxcount_k} into Eq.~\eqref{eq:neff_th_def} gives
\begin{equation}
n_{\mathrm{eff}}^{\mathrm{th}}(k)
=
n_0-\log_b C
-\frac{D_{\mathrm{eff}}}{\ln b}\ln k,
\label{eq:neff_th_intermediate}
\end{equation}
where $C$ is the proportionality constant in
Eq.~\eqref{eq:boxcount_k}. Imposing
$n_{\mathrm{eff}}^{\mathrm{th}}(1)=n_0$ gives $C=1$, and hence
\begin{equation}
n_{\mathrm{eff}}^{\mathrm{th}}(k)
=
n_0-\frac{D_{\mathrm{eff}}}{\ln b}\ln k.
\label{eq:neff_th_k}
\end{equation}
Equivalently, in terms of the normalized grid size,
\begin{equation}
n_{\mathrm{eff}}^{\mathrm{th}}
\left(\frac{\Delta}{H}\right)
=
n_0+\frac{D_{\mathrm{eff}}}{\ln b}
\ln\left(\frac{\Delta}{H}\right).
\label{eq:neff_th_dh}
\end{equation}

Because the finite fractal trees do not exhibit exact self-similarity
over all scales, $D_{\mathrm{eff}}$ is evaluated separately for each
neighboring grid-size interval in
$\Delta/H\in\{1,\,1/2,\,1/4,\,1/6,\,1/8,\,1/10\}$.
Within the interval $k\in[k_i,k_{i+1}]$, the prediction is evaluated
as
\begin{equation}
n_{\mathrm{eff}}^{\mathrm{th}}(k)
=
n_{\mathrm{eff}}^{\mathrm{th}}(k_i)
-
\frac{D_{\mathrm{eff},i}}{\ln b}
\ln\left(\frac{k}{k_i}\right),
\qquad
k\in[k_i,k_{i+1}],
\label{eq:neff_th_piecewise}
\end{equation}
where continuity is imposed by sequentially evaluating
$n_{\mathrm{eff}}^{\mathrm{th}}(k_{i+1})$ from
$n_{\mathrm{eff}}^{\mathrm{th}}(k_i)$.

\section*{Acknowledgments}
This work used computational resources TSUBAME4.0 supercomputer provided by Institute of Science Tokyo through Joint Usage/Research Center for Interdisciplinary Large-scale Information Infrastructures and High Performance Computing Infrastructure in Japan (Project ID: jh250082).

\bibliographystyle{elsarticle-harv}
\bibliography{references}

\end{document}